\documentclass[a4paper,10pt]{article}
\usepackage{graphicx}
\usepackage{longtable}
\title{The 3.5 kev line from non-perturbative
Standard Model dark matter
balls}
\author{C.D.~Froggatt${}^{1,2}$, H.B.~Nielsen${}^{2}$
\\[15mm] \itshape{${}^{1}$ Department of
Physics and Astronomy,}\\[3mm] \itshape{Glasgow University,
Glasgow, Scotland}\\[3mm] \itshape{${}^{2}$ The Niels Bohr
Institute, Copenhagen, Denmark}\\[3mm]
}
\begin{document}

\maketitle

\begin{abstract}
Our earlier put forward model of dark
matter, consisting of cm-size pearls with
ordinary matter inside under high
pressure and with
a mass of order $1.4 *10^8$ kg,
is used to explain the mysterious 3.5 keV
X-ray line from  the galactic
center and various galaxies and galaxy clusters.
The pearls are bubbles of a new type of
vacuum and thus surrounded by a surface
tension providing the high pressure.

We have two rather successful order of magnitude numerical
results:

1) the X-ray
energy of 3.5 keV
comes out as the homolumo-gap or rather
as the energy due to screening of
electrons in the high pressure
ordinary matter inside the pearls, and is
well fitted.

2) Using the fitting of Cline and Frey for
dark matter radiation arising from
collisions or annihilations of dark matter
particles we fit the overall  intensity
of the radiation in our pearl model.

We find that a pearl of the minimal size required just by stability, as used
in our previous work \cite{Tunguska}, is inconsistent with the observed
frequency and intensity of the 3.5 keV line. However the predictions of our
model are very sensitive to the radius of the pearls and an excellent fit to
both experimental quantities is obtained for a pearl of radius of 2.8 cm.

\end{abstract}

\section{Introduction}
We have for some time worked on a model
\cite{Dark1,Dark2,Tunguska,Supernova} for dark matter being balls
of cm-size (pearls) with mass of the order of $1.4 *10^8 kg \sim 100000$ tons,
consisting of ordinary matter highly compressed in a
bubble of a new type of vacuum called ``condensate vacuum''.
It is the purpose of the present article, using the
parameters of our earlier speculations and fits to
predict the intensity and frequency of an X-ray
line expected to be emitted from our dark matter pearls.
This emission is then of course to be identified with the controversial
3.5 keV X-ray radiation that somewhat mysteriously has been
observed by several satellites \cite{Bulbul,Boyarsky}.

Let us first list a series of remarkable characteristics
of our dark matter model:

\begin{itemize}
\item Contrary to most other proposals for what dark matter
could be, our model is in principle built into the Standard Model,
needing for its realization of the new vacuum only an
appropriate fine tuning of the coupling parameters of this
Standard Model according to our proposed ``Multiple
Point Principle'' \cite{MPP1,MPP2,MPP3,MPP4,tophiggs,Future}.
So no new physics needs to be found at LHC
in order that our model could be true.
\item We have proposed \cite{Tunguska} that the fall of one of our
dark matter pearls on the
earth is to be identified with the Tunguska event in 1908 in
Siberia.
\item And we suggest remnants from earlier
impacts of our pearls on the earth are to be identified with kimberlite pipes
\cite{Tunguska,Paszkowski,Corfu2015}, of which about
6500 have been found around the earth. Most kimberlite pipes
produced through the history of the earth must of course have
been covered by sediments through the earth history, and it may
thus not be unexpected that the still accessible kimberlite
pipes are found in old cratons, the very oldest sediments
on the earth.
\item An interesting property of our model,
assuming that  our pearls collect in the stars developing
into supernovae \cite{supernova}, is that they can cause the major
neutrino outburst from the supernova to {\em split into two outbursts},
as apparently occurred in the supernova SN1987A
\cite{MontBlanc,Kamiokande1,Kamiokande2,IMB,DeRujula}.
The idea
here is that our pearls, when confronted with a high density
of neutrons, absorb a lot of neutrons and heat up so
as to temporarily stop the collapse of the star. That heating
puts an end to the first major neutrino burst, but some several
hours later the heat production has stopped and enough heat
escaped by neutrino emission to let the collapse restart. Thereby
comes the second major neutrino burst.
\item Our pearls also help to make the supernova truly
become a supernova, in the sense of emitting visible material
to the outside of the star and thus becoming indeed visible from
earth as a spectacular new star.
\item According to our model of dark matter, the pearls form
in the first tenth of a second or so of the life time of the universe.
Close to the end of this process they have a fusion
caused explosion
transforming helium into heavier nuclei inside the pearls
and thereby emitting nucleons with the fusion energy.
This leads to an estimate \cite{Dark1,Dark2}
of the ratio of the amount of dark matter to normal matter outside the
pearls - meaning in practice ordinary matter - to be about 6,
in very good agreement with observations. The idea here is
that the nucleons
pushed out by the He-fusion caused
explosion is the matter which we today see as the ordinary
matter. What remains inside the dark matter pearls is also ordinary matter,
but that is not in practice observed as such today.
\end{itemize}

The observation of a new X-ray emission line at 3.5 keV was
reported \cite{Bulbul,Boyarsky} in 2014. It was detected in the
Andromeda galaxy, the Perseus cluster and different combinations
of other galaxy clusters. Later the line was detected in the
Galactic Center \cite{Boyarsky2,Jeltema,Carlson}.
This line has been suggested to originate from radiatively
decaying dark matter usually identified as  sterile
neutrinos \cite{Dodelson,Shi,Merle} of mass 7 keV.
However this interpretation of the data is controversial, as
the line was not observed in the blank sky (Milky Way halo)\footnote{
Recently there have been conflicting claims on observation: Boyarsky et al. \cite{Surface} see some 3.5 keV radiation in the
Milky Way Halo, while Dessert et al. do not see it \cite{Dessert}.} \cite{Boyarsky} or in dwarf spheroidal galaxies
\cite{Malyshev,Draco1,Draco2}.
So the expected distribution of the 3.5 keV line radiation
as simply coming from decaying dark matter and therefore
proportional to the amount of dark matter present
is not quite supported. Rather
a distribution as coming from interactions of two dark
matter particles giving an emission rate proportional to
the square of the dark matter density and velocity
dispersion - such as from annihilation \cite{Toma,Brdar},
inelastic scattering \cite{FreyCline} or, as
we propose in the present article, the collision of two
dark matter particles - could fit a bit better.

A feature that at first looks like killing the main
idea that this 3.5 keV radiation should come from the dark
matter is that the line shows up in the
Tycho supernova remnant \cite{Jeltema}.
However it actually supports our model
for dark matter, because we namely have dark matter pearls
that will radiate with this line 3.5 keV whenever they
are energized some way or another, as e.g. by the supernova
remnant.

In the present article we shall estimate/calculate {\em two}
numbers concerning the 3.5 keV radiation and compare the
results gotten from our model with the astronomical observations:
\begin{itemize}
\item We shall estimate the screening energy per electron
in the highly compressed material in the inside of our pearls,
i.e. the energy lowering achieved by the electrons around
a given electron adjusting - moving away appropriately -
to diminish the energy. We then expect that this screening
energy lowering functions like a homolumo gap effect \cite{Teller, Andric} and
causes there to be a gap of the size given by the screening
energy in the electron spectrum. Using this energy gap
excitons may be formed as weakly bound states of
an electron above the gap
and a hole below.
Such excitons are now supposed to decay under emission
of most of their energy as an X-ray and thus produce the
mysterious 3.5 keV line.

The major point of our first calculation is to show
that we order of magnitude-wise indeed get close to
3.5 keV.
\item The next calculation concerns the emission rate of the 3.5 keV
line expected from the dark matter pearls of ours.
In order to obtain just crudely enough radiation compared to the
observed rate, we need
 that our pearls collide with
each other, rather seldomly though, forming a double
as heavy pearl; there is then a very
strong release of energy, due to the fact that the surface tension area
around the two colliding pearls is bigger than that around the doubly as heavy pearl. The energy
due to contraction of the surface/skin is supposed to be
mainly radiated out as 3.5 keV radiation.
\end{itemize}

\subsection{Comments on our ``Tunguska'' model }
In a previous article \cite{Tunguska} we put forward the model
that dark matter is made up from macroscopic size pearls of mass
$M (=m_B)=1.4*10^8$ kg supposed to consist of a region of a new vacuum
- another vacuum phase - surrounded by a domain wall with a surface
tension $S$, of which the cubic root is $S^{\frac{1}{3}}$ = 28 GeV.
The only new physics compared to what we hope comes out of the Standard
Model by just an enormously ambitious calculation is our Multiple
Point Principle. For our purpose this principle tells us that the energy densities of the
two vacuum phases are the same. Under the pressure from the surface tension
ordinary matter like e.g. carbon or silicon,... is brought to the density
of the order of $\rho_B = 10^{14}$ $\frac{kg}{m^3}$. It is the electron gas, which
is highly degenerate, that keeps up the major part of this high pressure and
the electrons have relativistic speeds typically. The major difference to be felt
by the ordinary material inside the pearl is that the Higgs field
expectation value is supposed to be smaller inside the
pearl phase than outside. This in turn means that both nucleons and
electrons have slightly smaller masses inside a pearl than outside, and that thus
the matter is pressed into the pearl by a potential barrier at the surface.

We used the fact that about hundred years ago there was the famous Tunguska event
of trees falling over a 70 km large range in Tunguska in Siberia; a rather
mysterious impact in as far as no proper meteor was found in spite of the
rather large effects.
We identified this kind of impact with one of
our pearls hitting the earth with a rate of approximately $r_B \approx \frac{1}{200\ \hbox{years}}= 1.5 *10^{-8} \; s^{-1}$.

In the old article \cite{Tunguska} we made the theoretical approximation
that the pearls were all of just the size that places them on the borderline
of collapsing, because the pressure is just about to overcome the potential
barrier $\Delta V$ across the surface wall.


The parameter $\Delta V$, which denotes the potential difference per nucleon
in passing the domain wall, was crudely determined by assuming that the Higgs
field inside the pearl vacuum was about half of the Higgs field
expectation value in the outside (= ``present vacuum'' ).


Then in the old paper we wrongly assumed
additivity of quark mass contributions to the nucleon masses, and we derived
the value $\Delta V = 10  \pm 7 $ MeV. But now we have learned from a referee
that the interaction of a nucleon and the Higgs field is more complicated
and in fact is mainly due to virtual heavy quarks in the nucleon. In fact
H. Y. Cheng, and
C. W. Chiang \cite{gNN} calculate the Higgs coupling of a nucleon
- they get very similar values for the  neutron and proton - using
an effective lagrangian
${\cal L}(x)$ = $g_{\phi_H NN}\bar{\psi}(x) \phi_H(x) \psi(x)+...$
obtaining:
\begin{eqnarray}
g_{\phi_H NN}&=& 1.1 *10^{-3}.
\end{eqnarray}
Thus we obtain the corrected value
\begin{eqnarray}
\Delta V &=& 1.1*10^{-3} *123\ \hbox{GeV}= 135\ \hbox{MeV}
\hbox{ for $<\phi_H>_{cond} = 123$ GeV}.
\end{eqnarray}
In this paper we shall also consider the possibility that the
Higgs expectation value should be zero $<\phi_H> = 0$ in the
condensate phase. In this case we obtain
\begin{eqnarray}
\Delta V &=& 1.1*10^{-3}*246\ \hbox{GeV} = 270\ \hbox{MeV}
\hbox{ for $<\phi_H>_{cond}=0.$}.
\end{eqnarray}
Here $\psi$ is the nucleon field, and $<\phi_H>_{cond}$ is the
vacuum expectation value of the Higgs field $\phi_H$ in the condensate
vacuum (the one inside our pearls). (Of course the Higgs vacuum expectation
value in the present vacuum is given by $<\phi_H>_{present}=246$ GeV as is well-known
from the Fermi-constant.)

Using the Tunguska event rate $r_B$, the borderline stability hypothesis
and the old value $\Delta V =10\ \hbox{MeV}$ led to an estimate \cite{Tunguska} for the surface tension of
$S^{1/3} = 28\ \hbox{GeV}$.
It agreed well with the theoretical estimate of $S^{1/3} =16$ GeV, which was essentially
derived by assuming the vacuum phases were caused by electroweak physics so
that the tension came out to be of the order of the electroweak
energy scale.
With the corrected value $\Delta V = 135$ MeV, the surface tension is corrected by the
same factor of 13.5 giving $S^{1/3} = 380\ \hbox{GeV}$, which is still of the order
of the electroweak scale. Also the radius is decreased by this factor 13.5
from R = 0.67 cm to R = 0.05 cm and the density is increased from $1.0*10^{14}$ kg/m$^3$
to $2.5*10^{17}$ kg/m$^3$.

With these assumptions we obtained the parameters presented in Table
\ref{T}, where we show old and corrected values.

\begin{center}
\begin{table}[h]
\caption{The parameters of our model picture of the Tunguska particle
as a ball of a new type of vacuum with a bound state condensate, filled
with ordinary white dwarf-like matter and on the borderline of stability.}
\label{T}
    \begin{tabular}{ | l | l | l | l |}
\hline
    Time Interval of impacts & $r_B^{-1}$ & 200 years &\\
\hline
    Rate of impacts& $r_B$& $1.5 *10^{-10}\ \hbox{s}^{-1}$  &  \\ \hline
    Dark matter density in halo & $\rho_{halo}$ &0.3 GeV/cm$^3$ &\\ \hline
    Dark matter near solar system & $\approx 2\rho_{halo}$ &0.6 GeV/cm$^3$ &  \\ \hline
    Mass of the ball & $m_B$ & $1.4*10^8$ kg &  \\
    \hline
Estimated typical speed of ball& $v$& 160 km/s& \\ \hline
Kinetic energy of ball & $T_v$ & $1.8*10^{18}$ J&  430 megaton TNT \\\hline
Energy observed in Tunguska& $E_{Tunguska}$ & $(4-13)*10^{16}$ J  &
10-30 megaton TNT\\ \hline
 Potential shift between vacua& $\Delta V$ & 10 MeV $\rightarrow$
 135 MeV & \\ \hline
Cube root of tension & $S^{1/3}$&28 GeV $\rightarrow$ 380 GeV& from $m_B$ and rate\\ \hline
Cube root of tension& $S^{1/3}$ & 16 GeV & from condensate \\ \hline
Ball density& $\rho_B$ & $10^{14} \rightarrow 2.5*10^{17}$ kg/m$^3$ & \\ \hline
Radius of ball & $R$ & 0.67 cm $\rightarrow$ 0.05  cm& \\ \hline

\end{tabular}
\label{table}
\end{table}
\end{center}


The physics behind the two vacua is that the vacuum inside the
pearls, called the ``condensate vacuum'', results from bose condensation
of a speculated bound state of 6 top + 6 anti-top quarks. The scenario is that
the top-quark Yukawa coupling to the Higgs boson $g_t =0.935$ is actually
so large - taking into account the many colors and spin states of the
top-quark - that non-perturbative effects come in and e.g. cause the formation of a bound state
of the mentioned 6 top and 6 anti-top quarks \cite{gt}. We have six of each because this represents
a closed shell in the atomic physics sense. One shall imagine that these bound states
interact with each other, mainly by Higgs boson exchange, and bind to their neighbors
so as to make the energy (density) of the self-interacting condensate just
zero compared to the vacuum outside, called the ``present vacuum''. This is
supposed to come about by our proposed Multiple Point Principle, which
is supposed - in a mysterious way - to fine-tune the couplings in the Standard Model
so as to precisely achieve this kind of degeneracy between several vacua.

We introduce an effective field for the bound state,
which we may here call $F$, denoted $\phi_F$ and write an effective potential
as a function of both this field $\phi_F$ and of the Higgs field $\phi_H$.
Then the fine-tuning from the multiple point principle should organize the
Standard Model couplings in such a way that this effective potential $V_{eff}(\phi_H, \phi_F)$
has two different and equally deep minima corresponding to the two degenerate
vacua.
In our previous work \cite{Tunguska} we assumed that, while in the ``present vacuum'' the
$\phi_F$ field has zero expectation value, both fields $\phi_F$ and $\phi_H$ have
non-zero expectation values in the ``condensate vacuum''.
We then took as a crude guess that
the Higgs expectation value
in the ``condensate vacuum'' $<\phi_H> = 123$ GeV equal, as already mentioned, to half the value
of the one in the ``present vacuum''.

However we present a theoretical investigation of the effective potential $V_{eff}(\phi_H, \phi_F)$
in the appendix, which suggests that in the ``condensate vacuum'' it is more natural for the
Higgs field expectation value to be zero $<\phi_H> = 0$, somewhat in analogy to the assumption
that the expectation value of the $\phi_F$ field is zero
in the ``present vacuum''.
Thereby we get a bigger difference between
the
interactions of the nucleon in the two phases and thus a larger value, by a factor of 2, for the mass or potential
energy difference $\Delta V$ between the two phases for a nucleon. Therefore
in the present paper we prefer to take the value
$\Delta V = 270 $ MeV
rather than $\Delta V = 135  $ MeV.

Another modification or improvement of our model in the present article
compared to the original one \cite{Tunguska}, is that we now believe it
to be unrealistic to assume that the size of the pearl is just so that it is
on the stability borderline, because
a tiny little vibration - especially during its formation in the big bang era - would have caused
it to collapse. Rather we introduce in section \ref{size} a parameter $\xi$ denoting the ratio of the
actual (average) radius $R$ of the pearls relative to the critical radius $R_{crit}$ used in \cite{Tunguska}
corresponding to the pearls being in the critical state just about to collapse.


Throughout the paper we shall present results using both the parameters in
Table \ref{table} and with the above improvements.

\section{The Frequency of the 3.5 keV X-ray}
\label{s2}
\subsection{Estimate of Energy Gap}

Now we shall estimate the homolumo energy gap \cite{Teller, Andric}, which is supposed to appear between the
filled and the empty states in the material in the interior of our dark
matter pearls.
Basically we shall first argue in subsection \ref{existence} that there will be
a homolumo gap in the supposed fluid or glassy state of the highly compressed
material inside our
pearls. Next we shall give a very simple dimensional argument in subsection
\ref{dimensionarg}, that the order of
magnitude of this gap for our relativistically compressed matter shall
be given in terms of the Fermi energy/momentum $E_f = cp_f$ by

\begin{equation}
 \hbox{``homolumo-gap''} \approx \alpha^{3/2}c^{-1/2} p_f \approx
\alpha^{3/2} c^{-3/2} E_f.
\end{equation}

\subsection{Existence of Energy Gap}
\label{existence}

It is not easy to know what sort of state, such as crystalline or fluid,
that ordinary matter should take under the enormous pressure inside our pearls.
So we shall allow ourselves to speculate, that it is in a fluid
or glassy state in which there is a lot of irregularities.

We shall assume that in such a glass-like material it is
typical to find a gap in the electron state spectrum.

If the glassy structure is very pronounced the electron
energy eigenstates tend to be (Anderson) localized, and
effectively the spectrum should be determined from a local
region in the material.
A priori, before one takes into account the back reaction
from the interaction
between the electrons with themselves or other degrees of
freedom, the electron spectrum would be given by the
band structure or let us say better by some random matrix
model. In that situation all the intervals between neighboring
energy levels will be similar in magnitude. But now, when
back reactions from the electrons in the filled states are
taken into account, the special level spacing between the highest
occupied homo- and the lowest unoccupied lumo-levels  would
be expanded to be much larger than the other spacings.
It is this expansion we call the homolumo-gap-effect \cite{Andric}.

In the present article we want to assume that
for some reason there is an expanded (due to the interaction)
homolumo gap. Then below we
shall seek to estimate the size of this homolumo gap
from the philosophy that it appears due to the interaction
that an electron has with neighboring electrons, which
consequently adjust to minimize the total energy in the
presence of the first considered electron.
The interaction between the electrons
and other charged particles - the protons or the electrons
themselves - is basically due to the Coulomb force and thus
we are led to estimate the homolumo
gap effect from such a Coulomb interaction. This means that
the homolumo gap effect is essentially the same as the screening
of the electric charge of the electrons in the filled states.
We therefore suggest in the present article to estimate the
homolumo gap effect by using an estimate of the screening based
on the Thomas Fermi approximation \cite{Mermin}.

\subsection{Dimensional Estimate of Energy Gap}
\label{dimensionarg}

We shall here use units with Planck's constant $\hbar = 1$.
Note however, that we do NOT here put the light velocity to 1 as one would
normally have done in
high energy physics.
With the Planck constant $\hbar$ put to 1, it follows that
the distance $r$ multiplied by a
momentum $p$ is dimensionless $[rp]=[\hbar]=1$.
Since $\alpha/r$ has the units of energy, it then follows that
the fine structure constant $\alpha$ has the dimension of
$J[r] =J/[p]=kg [v^2]/kg [v]=[v]$,
where $v$ is a velocity.

However we consider such a high density of matter that, because of the pressure,
the electrons  become relativistic. Consequently
{\em two} velocities appear in the calculation,
namely both the fine structure constant $\alpha$ and the light velocity $c$.

In order to make a dimensional argument for the homolumo gap in this
relativistic case, we shall
extract the main dependence on the speed of light from the Thomas Fermi
based calculation
in section \ref{ThomasFermi}. The crucial quantity for calculating the
screening energy - which
we identify energy-wise with the homolumo gap - is the derivative
of the density of electrons $n_e$  with respect to the Fermi energy $E_f$ of
the electrons in the material.
In the units with Planck's constant put to unity the density is given as
$p_f^3$ dimensionally/order of magnitudewise.
The (Fermi) energy of an electron is $E_f = cp_f$, and thus the derivative
of the density of
electrons w.r.t. the Fermi energy $\frac{\partial n_e}{\partial E_f}$
is given dimensionally by $p_f^3/(p_fc) =p_f^2/c$. In addition to this
derivative we need no
more information about the sea of electrons to calculate its screening
effect, and thus the homolumo gap
will only depend on this derivative $p_f^2/c$ and on the fine structure
constant $\alpha$,
which occurs in the driving force for screening. The homolumo-gap or the
screening energy  of course has
dimension of energy [J], and thus in our units for now
we have {\em ignoring dimensionless factors}
\begin{eqnarray}
 E_H &=& \sqrt{\frac{p_f^2}{c}} \alpha^{3/2} \hbox{ (order of magnitudewise).}\nonumber\\
&=& p_f\sqrt{\frac{\alpha^3}{c}} = E_f \sqrt{\frac{\alpha^3}{c^3}}
\label{EH}
\end{eqnarray}

As in \cite{Tunguska} we at first assume that our pearls are on the
borderline of stability, which leads to the following result
for the Fermi energy of the relativistic degenerate electrons
\begin{equation}
 E_f = 2 \Delta V =270 \ \hbox{MeV}\quad  \hbox{for $\phi_{H}=$ 123 GeV}. \label{Ef}
\end{equation}
or
\begin{equation}
 E_f = 2 \Delta V = 540 \ \hbox{MeV}\quad  \hbox{for $\phi_{H}=0$}. \label{Ef2}
\end{equation}


  Ignoring factors of order unity and thus only making a dimensional argument
we get using (\ref{EH}) and (\ref{Ef})
\begin{eqnarray}
 E_H &=& \sqrt{\frac{\alpha^3}{c^3}}270\ \hbox{MeV} \\
&=& 170 \ \hbox{keV} \quad  \hbox{for $\phi_{H}=$ 123 GeV}.
\end{eqnarray}
or
\begin{equation}
E_H = 340 \ \hbox{keV} \quad  \hbox{for $\phi_{H}=$ 0}.
\end{equation}
This is too large by about a factor 50 or 100, but the uncertainty due to the unknown
radius for example
is high.


\subsection{Thomas Fermi Calculation}
\label{ThomasFermi}

Now we want to estimate the order of unity factors ignored in the above
dimensional argument. For this purpose we shall use the semiclassical
Thomas Fermi approximation \cite{Mermin}. In this approximation we
estimate the effect of an electric field, say, using macroscopic
considerations of Fermi surface statistical mechanics over
infinitesimally small pieces of matter.


Our purpose is to calculate the homolumo gap expected to occur as a gap in the
single electron spectrum due to back reaction from the electrons themselves.
We plan to do that by calculating the decrease in energy  of the system around an inserted
charged particle (e.g. an electron) due to screening. That is to say we consider the decrease in
energy as being due to the displacement of the charged particles in the medium adjusting
to the inserted charged particle, and that is the screening. The potential around the inserted
charged particle is of course $\alpha/r$ where $r$ is the distance to this inserted charge.
The important thing to estimate is thus the screening-length
for this inserted charge.
The potential really means
that at the distance $r$ the Fermi-energy gets shifted by the amount $\alpha/r$, and thus
the  density of the charged particles in the medium (the electrons) get changed by
$\alpha/r * \frac{\partial n_e}{\partial E_f}$. The derivative $\frac{\partial n_e}{\partial E_f}$
of the electron density
\begin{eqnarray}
 n_e &=& 2 \frac{1}{(2\pi)^3}*\frac{4}{3}*\pi p_f^3 = \frac{1}{3\pi^2} p_f^3
 \label{ne}
\end{eqnarray}
w.r.t. the Fermi energy $E_f = c p_f $ in the ultra relativistic limit
of the electron moving almost with speed of light, becomes
\begin{eqnarray}
 \frac{\partial n_e}{\partial E_f}
&=& \frac{p_f^2}{\pi^2 c}.
\end{eqnarray}

 Close to the inserted particle, i.e. for small $r$, the change in the density due to
the screening is simply  $\frac{\alpha}{r} * \frac{\partial n_e}{\partial E_f}$, but as
we go to large $r$ part of the inserted charge has already been screened and thus we
should rather use a diminished charge $\approx 1-\int_0^r 4\pi \frac{\alpha}{r} \frac{\partial n_e}{\partial E_f}* r^2 dr$.
More precisely we shall calculate the unscreened charge $Q(r)$ measured in Millikan charge quanta  at a distance $r$ from the
inserted point charge using the differential equation
\begin{eqnarray}
 \frac{dQ(r)}{dr} &=& -4\pi r^2  Q(r)\frac{\alpha \frac{\partial n_e}{\partial E_f}}{r}\\
&=& - k_0^2 rQ(r)\\
\hbox{where} \quad k_0^2 &=&  \frac{4\alpha p_f^2}{\pi c} \quad  \hbox{and} \quad Q(0) = 1\label{f13}
\end{eqnarray}
It is easy to see that the solution of the differential equation is given
by the Gaussian
\begin{eqnarray}
Q(r) &=& \exp(-\frac{k_0^2}{2} * r^2)
\end{eqnarray}
The screening energy is thus in first approximation given by an integral over the energy per charge $\alpha/r$ multiplied by the induced
charge density distribution $\frac{\alpha Q(r)}{r}  *\frac{\partial n_e}{\partial E_f}4\pi r^2 dr$ and thus becomes
\begin{eqnarray}
 E_{HW}&=& \int_0^{\infty}\frac{\alpha}{r} * \frac{\alpha Q(r)}{r}  *\frac{\partial n_e}{\partial E_f}4\pi r^2 dr\\
&=& \alpha k_0^2 \int_0^{\infty} \exp(-\frac{k_0^2}{2} r^2)dr\\
&=& \alpha k_0 \sqrt{\frac{\pi}{2}}\\
&=& \sqrt{2}\alpha^{3/2}c^{-1/2} p_f
\end{eqnarray}

\subsubsection{Energy of medium due
to the screening change}

The quantity $E_{HW}$ calculated just above was the energy decrease of the Coulomb interaction of the
inserted charge with the charges in the medium being pushed somewhat away in the screening.
We, however, did not yet take into account that these particles in the medium would thereby increase
their energy in their interaction with each other or something else than just this
inserted particle. Thus the above calculated decrease in energy has to be diminished
by this increase in medium energy caused by the screening disturbing the medium.
We have estimated the effect of this disturbance by imagining that the electrons
in the medium function like harmonic oscillators and that the charge of the inserted
particle is gradually increased to its actual value. In this way we find that the
disturbance increases the energy of the medium by just $E_{HW}/2$. So overall our
estimate of the decrease in energy due to the screening is
\begin{equation}
 E_{decrease} = E_{HW} - E_{HW}/2 =E_{HW}/2.
\end{equation}

Now, however, what we are really interested in is the homolumo gap, in the sense
of asking for the energy increase when an electron is moved from a homo-state to
a lumo-state. The moved electron then leaves behind a hole with an anti-screening
energy of the opposite sign $E^{hole}_{decrease} = -E_{HW}/2$, due to the interaction energy
of the electrons surrounding the hole.
Hence the homolumo gap ends up being just
\begin{equation}
E_H = E_{decrease} - E^{hole}_{decrease} = E_{HW} =
\sqrt{2}\left(\frac{\alpha}{c}\right)^{3/2} E_f.\label{EHW}
\end{equation}


Inserting
$E_f =$ 270 MeV or 540 MeV
and $\alpha/c = 1/137$ into (\ref{EHW}),
we obtain for our estimated homolumo gap
\begin{eqnarray}
 E_H &=& 270\ \hbox{MeV}* 137^{-3/2} \sqrt{2}\\
&=&  240 \ \hbox{keV}\quad} \hbox{(for $\phi_H= 123\, GeV$ ) \label{EHold1}
\end{eqnarray}
or
\begin{eqnarray}
E_H &=& 480\ \hbox{keV} \quad \hbox{(for $\phi_H=0 $ )} \label{EHold2}
\end{eqnarray}
Again this result is two orders of magnitude too high.


\section{Line Intensity}
The model proposed for how the 3.5 keV line gets energized is that pairs of our pearls
meet each other, very seldomly of course.
In the collision of two pearls  the skin surrounding the two pearls contract to only one skin
surrounding a united bubble. But now since the skin has a very high tension
the energy released by this skin contraction is a few percent
of the Einstein energy of a whole pearl.


\subsection{Ratio of Gain of Energy
by Pearl Collapse to Mass}
In this subsection we shall calculate the ratio of the energy $E_R$ released, when two
of our pearls unite to the mass of one such pearl $M$:

We shall do that calculation of the
ratio $\frac{E_R}{Mc^2}$ in terms of the
fermi-momentum $p_f$ of
the degenerate electrons in the pearl.
This Fermi-momentum cannot be larger
than $2\Delta V$,
because if so the electrons would
pull the nucleons
out of the pearl \cite{Tunguska},
\begin{equation}
 p_f \le 2\Delta V.\label{pfDv}
\end{equation}
However, we shall at first
assume  - as in our earlier paper
\cite{Tunguska} -  that the pearls are
just on the borderline of stability, so
that we indeed have equality in the above
inequality (\ref{pfDv}).
We take it that there are 2 nucleons per
electron, but really we
expect the ordinary matter in the pearl
to be slightly over rich in neutrons so
this 2 might go up a bit. Thus, using
(\ref{ne}) for the electron density,
the mass-density $\rho_B$ of pearls
becomes
\begin{eqnarray}
 \rho_B &=& 2m_N *n_e\
=\ \frac{2m_Np_f^3}{3\pi^2}.
\end{eqnarray}
The radius $R$ of a pearl is
related to its mass by $M =\frac{4\pi}{3}R^3\rho_B$
and hence
\begin{eqnarray}
R &=& \frac{1}{p_f}\left(
\frac{9\pi M}{8m_N}\right)^{1/3}. \label{R}
\end{eqnarray}

When a collision occurs between two
dark matter balls, an energy $E_R$ of
the order of the energy in the bubble surface
$E_S$ of a single pearl is released.
Denoting the surface tension, the
tension  of the skin around the pearl,
by $S$ the pressure from this skin is
$2S/R$. This pressure must equal the
pressure from the relativistic electrons,
which dominates the pressure of the
material in the pearl,
 \begin{eqnarray}
  \frac{2S}{R} &=& P\ = \ \frac{c p_f^4}
{12\pi^2}. \label{pressure}
 \end{eqnarray}
Thus the energy of the skin, the surface
energy, is
\begin{eqnarray}
 E_S &=& S * 4\pi R^2 \ =\  R^3\frac{c p_f^4}{6\pi}\\
&=& \frac{3}{16}\frac{Mc p_f}{m_N} \label{ES1}
\end{eqnarray}

So the fraction of the Einstein energy
$Mc^2$ of a pearl
that is emitted in some way, which we
suggest to be mainly in the form of the
3.5 keV X-ray line, when two particles
collide, is
\begin{eqnarray}
 \frac{E_R}{Mc^2} \sim \frac{E_S }{Mc^2}
&=& \frac{3}{16} * \frac{cp_f}{m_Nc^2}
\label{ESoverM}
\end{eqnarray}

With $p_f = 2\Delta V =
270
\ \hbox{MeV/c} $ and $m_Nc^2=940\ \hbox{MeV}$, we thus find
\begin{eqnarray}
 \frac{E_S}{Mc^2}
 &=& \frac{3}{16} *
0.27\\
&=&
5.4\%
\end{eqnarray}
With $p_f = 2\Delta V =
540\ \hbox{MeV/c} $ (as for $\phi_H=0$) we get
\begin{eqnarray}
 \frac{E_S}{Mc^2}
&=&
10.8 \%.
\end{eqnarray}


\section{Fit by $\rho^2$, Cline Frey}

\subsection{Cline Frey work}

In the article \cite{FreyCline} Cline and Frey fit the observations of the
3.5 keV line from the point of view
that the
line is due to inelastic scattering of dark matter to an excited
state that subsequently decays - the
mechanism of excited dark matter (XDM). The speciality of such an
XDM model is that the intensity of radiation coming from a
region in space is proportional to {\bf the square of the density
$\rho_{DM}$ i.e. to $\rho_{DM}^2$ rather than to the first power only}.
In a model like ours, in which the production of the X-ray line 3.5 keV
comes from collisions of our dark matter pearls, of course
the production rate of these X-rays also goes proportionally to the square
of the density.

\subsection{Their analysis}

\begin{table}[h!]
\caption{This table is based on the table 1 in reference \cite{FreyCline}.}
\begin{tabular}{|c|c|c|c|c|c|}
\hline
Name&$N<\sigma_{CF} v>*$&$v$&boost&$(\frac{N<\sigma_{CF} v>}{v *boost})*$&Remark\\
&$\left ( \frac{10GeV}{M} \right )^2$&&&$\left ( \frac{10GeV}{M} \right )^2$
&\\
Units&$10^{-22} cm^3s^{-1}$&$km/s$&&$10^{-27}cm^2$&\\
\hline
Clusters\cite{Bulbul}&480 $\pm$ 250 & 975& 30& 0.016 $\pm$ 0.008&\\
Perseus\cite{Bulbul}& 1400 - 3400&1280& 30& 0.037 - 0.09&\\
Perseus\cite{Boyarsky}& (1 - 2) $*10^5$&1280&30&2.7 - 5.3&ignored\\
Perseus\cite{Urban}&2600 - 4100& 1280& 30& 0.07 - 0.11&\\
CCO\cite{Bulbul}&1200 - 2000& 926& 30&0.04 - 0.07&\\
M31\cite{Boyarsky}&10 - 30(NFW)&116&10& 0.0086 - 0.026 &\\
&30 -50 (Burkert)&&&0.026 -0.043&\\
MW\cite{Boyarsky2}&0.1 -0.7 (NFW)&118&5& 0.00017 - 0,0012&ignored\\
&50 -550 (Burkert)&&&0.084 - 0.93&in average\label{table0}\\
\hline
Average&&&&0.032$\pm$ 0.006&\\
\hline
\end{tabular}
\end{table}
The main result of the analysis of Cline and Frey is reproduced by their
table given just above.
The notation used is as follows:
\begin{itemize}
\item N is the number of 3.5 keV photons emitted per collision.
\item $\sigma_{CF}$ is the cross section as calculated assuming that there is
no boost (see below) for the particles of the dark matter
to collide. That is to say $\sigma_{CF}$ is calculated as if the density
$\rho$ of dark matter were given just by the model
for dark matter distribution used by Cline and Frey, {\bf without any
clumping further than the galaxies themselves having been included}.
The estimated true cross section is thus rather
\begin{equation}
 \sigma = \sigma_{CF}/boost.\label{sigma}
\end{equation}

\item $v$ is the average velocity dispersion.
\item $boost$ is the increase in collision rate due to the clumping
of the dark matter into sub-halos, which we crudely estimate from
references \cite{Moline}, \cite{Okoli} and \cite{Hiroshima}.
These papers contain computer simulations of gravitational interactions
between dark matter constituents in galactic halos forming clumps of much higher density than the average density.
Such clumping obviously enhances the true average square density
compared to the square of the average density.
\item $M$ is the mass of the dark matter particles.
\item Since it looks hopeless to get a fit
with
values for the same quantity differing by a factor 50 or 100, we left out the Boyarsky measurement for the
Perseus cluster. It really means that the model used by Cline and Frey with the
intensity proportional to the {\bf square} of the dark matter density
is disfavored by the Perseus cluster analysis. In the section
\ref{Perseus} below we mention other troubles for precisely the Perseus Cluster,
possibly supporting the idea that something there goes on which we do not understand well.
\item Because of the uncertainty in the dark matter distribution in the
Milky Way center (MW) - whether it be NFW or Burkert - we left the Milky Way center   out of the averaging;
but within uncertainties it is quite consistent with the average.
\item The notations ``NFW''\cite{NFW} and ``Burkert''\cite{Burkert} stand for a couple of different
models for the distribution of dark matter in a galaxy mainly deviating
by ``NFW'' having a strong peak at the center of the galaxy.
\item MW stands for our Milky Way galactic center.
\item M31 is the Andromeda Galaxy.
\item CCO stands for a combination of Coma + Centaurus + Ophiuchus clusters.
\end{itemize}

 Using the average value for
$(\frac{N<\sigma_{CF} v>}{v *boost})*\left ( \frac{10\ GeV}{M} \right )^2$
from the table and (\ref{sigma}), we obtain
\begin{eqnarray}
 \left(\frac{N\sigma}{M^2}\right)_{exp} &=& (0.032 \pm 0.006)  * 10^{-27}cm^2 /(10\ GeV)^2\\
&=& (1.0 \pm 0.2)*10^{23}cm^2/kg^2\label{Freyf}
\end{eqnarray}

\subsection{Trouble of Perseus Cluster}
\label{Perseus}
The amount of 3.5 keV radiation from the Perseus Cluster is controversial:
At first it seemed to be a significant source \cite{Bulbul,Boyarsky}
suggesting a lifetime of the order of
$3*10^{27}s $ in the sterile neutrino model.
But then the Hitomi satellite \cite{Hitomi} did not see any 3.5 keV signal
from the Perseus Cluster.
Now a possible way out of the
controversy would be \cite{absorption} that there is
a 3.5 keV absorption line which for Hitomi (which had less angular resolution
and thereby included the active galactic nucleus
in their observations) would compensate the diffuse cluster emission line. Such a story about an
absorption line is, however, totally unacceptable in our model.
Our pearls would certainly not
be able to absorb radiation of any significance from Perseus.

Now, however, this absorption picture has severe problems by
itself: In fact Conlon et al. \cite{absorption}
have suggested a fluorescent dark matter model
to solve the Perseus Cluster problem, in which a 3.5 keV
absorption line results from resonant excitation of a dark
matter particle $\chi_1$ of mass $m_{DM}$ much greater than 3.5 keV.
The excited dark matter particle $\chi_2$ then drops back
to its ground state $\chi_1$, providing the 3.5 keV emission line
seen in the diffuse cluster.
Making an ansatz for the fluorescent dark matter particle
interaction with the photon of the form
\begin{eqnarray}
 {\cal L} \supset \frac{1}{M} \bar{\chi_2} \sigma_{\mu\nu}\chi_1F^{\mu\nu}.
\end{eqnarray}
(In this subsection $M$ is the inverse of the effective coupling.)
Conlon et al derive a lower bound for the $\chi_2 \rightarrow \chi_1 + \gamma$
decay width
\begin{equation}
 \Gamma \ge \left ( \frac{m_{DM}}{\hbox{GeV}}\right ) * (1 \hbox{ to }10)*10^{-10}\ \hbox{keV}.
\end{equation}
This relation leads to
\begin{equation}
 \frac{m_{DM}}{\eta^{2/3}}
\stackrel{<}{\sim}
10^6\ \hbox{keV},\label{ineq}
\end{equation}
where $\eta = m_{DM}/M$.

But now earlier Profumo and Sigurdson, see fig 2 in their paper
\cite{Sigurdson}, had investigated the experimental constraints
on the parameters $m_{DM}$ and $\eta$ in such a resonant absorption
model of dark matter. The region corresponding to (\ref{ineq}) in their
figure is incompatible with the allowed range for the
parameters. Thus the fluorescent model does not seem tenable.

Taking the intensity of the 3.5 keV radiation to be proportional to the square
of the dark matter density (as in the analysis of Cline and Frey),
the data at larger angles by Boyarsky et al. \cite{Boyarsky} are far too high compared to the data
from Bulbul et al. \cite{Bulbul} at lower angles from the direction to the center of the Perseus cluster.
This could mean a severe disfavoring of square density models.
In order to avoid having this discrepancy entering into our fit,
we simply leave out the Perseus results of Boyarsky et al., because it is
the Bulbul et al. results which agree best with the other data in the table.

\subsection{Prediction}

We found (see (\ref{ES1}) and (\ref{R})):\underline{}
\begin{eqnarray}
 E_S&=&\frac{3}{16}\frac{M p_f}{m_N} \\
R&=&\frac{1}{p_f}\left ( \frac{9\pi M}{8m_N}\right )^{1/3}
\end{eqnarray}
(we use c=1 units, except in sections \ref{s2} and \ref{s5}.)
and so
\begin{equation}
\sigma = \pi (2R)^2
= 4\pi* \frac{1}{p_f^2}\left ( \frac{9\pi M}{8m_N}\right )^{2/3}.
\end{equation}
Assuming 100\% efficiency in converting the energy released by the
surface contraction into the 3.5 keV radiation, we obtain
\begin{equation}
 N = E_S/(3.5\ keV)
= \frac{3}{16}\frac{M p_f}{m_N (3.5\ keV)}
\end{equation}
and, using the parameters $M=1.4*10^8$ kg and
$p_f=270$ MeV,
we find
\begin{eqnarray}
\frac{N\sigma}{M^2}& = & \frac{3}{16}\frac{M p_f}{m_N (3.5\ keV)}  *
\frac{4\pi}{p_f^2}\left ( \frac{9\pi M}{8m_N}\right )^{2/3}\frac{1}{M^2}\\
&=& 5.5\left(\frac{m_N}{M}\right)^{1/3}m_N^{-2}\ (p_f*3.5\ keV)^{-1}\\
&=&
2.0 *10^{21} cm^2/kg^2, \label{prediction}
\end{eqnarray}
for which the above data analysis gave
 $(1.0 \pm 0.2)*10^{23}cm^2/kg^2$.

This means we predict a factor of
fifty too little
production of 3.5 keV radiation for a critical size ball just on the stability
borderline. With our accuracy this prediction is already promising given that
the value is sensitive to the radius R. However we note that the result (\ref{prediction})
was made assuming that all the released energy went into
the 3.5 keV X-ray radiation. In fact we want to argue that this is
likely to be true in order of magnitude, but that is far from obvious at first.

\section{Efficiency of Sending
Energy to the 3.5 keV line}
\label{s5}


The heat energy of the electrons \cite{Mermin} in the pearl at a temperature T would be
\begin{eqnarray}
 \hbox{``heat energy (electrons)''}
&=& \frac{\pi^2T^2M}{4cp_f2m_N}
\end{eqnarray}
 and the  heat energy of the nucleons which is essentially
that of the phonons would be
\begin{eqnarray}
 \hbox{``heat energy (nuclei)''} &=& \frac{3TM}{m_N} \hbox{ (above the Debye temperature)}.
\end{eqnarray}
If the nucleons dominate the heat energy after the collision we have
\begin{equation}
\hbox{``heat energy (nuclei)''} = E_S.
\end{equation}
Using the estimate (\ref{ES1}) $E_S = \frac{3Mcp_f}{16m_N}$ for the energy
released by the contraction of the skins of two colliding pearls,
we then obtain the following equation for the temperature after the collision
\begin{equation}
 \frac{3TM}{m_N} = \frac{3Mc p_f}{16m_N}
\end{equation}
or
\begin{equation}
T = \frac{cp_f}{16}.
\end{equation}
 In the case the electrons should have dominated the heat energy we would have got
 \begin{equation}
 \frac{T^2M\pi^2}{4p_fc2m_N} =  \frac{3Mcp_f}{16m_N}
 \end{equation}
 or
 \begin{equation}
 T = \sqrt{3/2} \frac{cp_f}{\pi}.
 \end{equation}

If we take into account that the nucleons are not oscillating separately,
because they are bound into nuclei of say $N$ nucleons
per nucleus, the temperature  which we found at first to be
$T=\frac{cp_f}{16}$ is replaced by
\begin{eqnarray}
 T &=& \frac{N cp_f}{16}\\
&\approx & \frac{3}{4} cp_f \hbox{ ( for C, Ne say, assuming C dominates )}.
\end{eqnarray}
But if so then the take up of heat by the electrons and the nuclei
become similar. Thus we must
add their contributions and the temperature will be about
the half of each of them separately. In fact combining the two heat capacity
sources, electrons and nuclei, we get (via a second order equation):
\begin{eqnarray}
 T &\approx &0.3 cp_f
= 0.6 \Delta V.
\end{eqnarray}
This $T$ is thus the
temperature
of the pearl 
if the energy from the
contraction of the surface has spread throughout
the pearl. We shall consider it to be the
temperature in the central region during
the cooling off of the pearl, suggested by us
to go by emission of the 3.5 keV radiation.
For the critical size pearl under consideration
\begin{eqnarray}
T &=&
81 \ \hbox{MeV} \quad  \hbox{if $\phi_{H \, condensate}=123$ GeV}\label{temp1}\\
&=&
162\ \hbox{MeV} \quad \hbox{if $\phi_{H \, condensate}=0$}\label{temp2}
\end{eqnarray}
and the temperature would be so high that the nuclei would dissociate into nucleons.

However this critical size is physically unrealistic and the actual pearl radius
must be larger than the critical radius $R_{crit}$. In section 7 we consider a pearl of radius R
and then fit the experimental data and theoretical constraints using the parameter
$\xi = \frac{R}{R_{crit}}$. We obtain fits for a range of $\xi$ values. A typical value
given later in Table 3 is $\xi = 23$ for fixed $\Delta V = 135$ MeV. Our
favoured fit to just the experimental data  actually gives $\xi = 57$.
The Fermi momentum $p_f$ is inversely proportional to $\xi$ and so the temperature T
of the pearl is reduced from 81 MeV to 3.5 MeV or 1.4 MeV for the typical
and favoured values of $\xi$ respectively. Thus in practice we expect that the dissociation of the
nuclei is a relatively small effect. We shall therefore ignore the dissociation of the nuclei, although
strictly incorrect to do so for a critical size pearl with $\xi = 1$ which we continue to formally use here.

But now the problem is whether the pearl material
has sufficiently low heat conductivity to
allow the surface to maintain a temperature
just of the order of magnitude of 3.5 keV
with such an enormously hot center.

\subsection{The heat conductivity.}
In this subsection we shall by dimensional
analysis estimate the heat conductivity $k$
for our pearl material as well as in parallel
for ordinary matter.

We shall assume that our material is in the high temperature
range which corresponds to what in ordinary metals
leads to a temperature independent conductivity $k$.
This means that we take it that the temperature is
higher than the Debye temperature, as discussed below.

The velocity of the sound is crudely estimated
as the square root of the pressure over the
density,
\begin{eqnarray}
 \hbox{``sound velocity''} &\approx& \sqrt{\frac{``pressure''}{``density''}}\\
&\approx & \sqrt{ \frac{c p_f^4/(12\pi^2)}{2m_Np_f^3/(3 \pi^2)}}
= \sqrt{\frac{c p_f}{8m_N}}\\
&=&
0.2 \label{sound}
\end{eqnarray}
Here we took $p_f = 270$ MeV/c.
The effective lattice constants would be
$p_f^{-1}$ if we took it that there were equally
many nuclei as electrons. But first there are both
a proton and a neutron for each electron, and secondly
these nucleons are collected into nuclei. So, taking N = 12
nucleons per nucleus,
the effective lattice constant becomes
\begin{eqnarray}
 \hbox{``effective lattice constant''} &=& p_f^{-1}\sqrt[3]{N/2}\\
&\approx& 0.01 c/\hbox{MeV}.
\end{eqnarray}

Thus the Debye frequency becomes of the order
\begin{eqnarray}
\hbox{``Debye frequency''} & \approx &\frac{\hbox{``sound velocity''}}{ \hbox{``effective lattice constant''}}\\
&=&
25\ \hbox{MeV}
 \end{eqnarray}

The temperature in the interior in the beginning  $\sim
81\ \hbox{MeV} $ (\ref{temp1})
 is significantly higher than this
Debye temperature of
25 MeV. So we are in the high
temperature regime where the heat conduction should be roughly temperature
independent. This is also true for a pearl with a radius larger than $R_{crit}$.


\subsubsection{The power of  $\alpha$ in the pearl matter conductivity.}

A priori one thinks that the mean free path is determined
from interactions and that each time we have an
interaction in a material this interaction must be
electromagnetic. Thus such interactions will always
be proportional to the electromagnetic coupling constant
$\alpha$, which in our units here is a velocity.

However when we consider phonons or equivalently
interactions with the vibrating nuclei, there is
an interesting cancelation preventing the coupling constant
$\alpha$ from coming into the calculation, except as the
electron velocity in the non-relativistic materials:

The point is that the vibration of the whole crystal
- or better above the Debye temperature  the vibration of the atomic nuclei -
is only driven back to its equilibrium state by forces that
are themselves proportional to the fine structure constant
$\alpha$. This means that the back driving force is the weaker
the smaller this constant $\alpha$. Thus the displacement of the
nuclei or equivalently the phonon amplitudes are the bigger the smaller is $\alpha$.

For instance a nucleus being displaced a distance $r$ from its
equilibrium position feels a potential proportional to $\alpha$.
Strictly speaking $\alpha/r $ say, but we could approximate it
by an harmonic potential $\alpha*r^2$ corrected by some constant of the order
of the ``lattice constant'' $a$ say to get the right dimension; then
the potential energy is approximated as $\alpha *r^2/a^3$.
In any case the distance the atomic nucleus gets displaced
by its thermal motion will be so as to increase its potential
energy by of order the temperature $\sim T$. With the harmonic
approximation the distance squared for its typical deviation then
becomes $r^2 \sim Ta^3/\alpha$. Now the electron being
stopped by interacting with such a nucleus has an interaction
that must be proportional to $\alpha$. So it gets an amplitude for
its scattering proportional to $r\alpha$, provided the
screened field of the electron hits the oscillating atom.
Most of the scattering of the electrons by phonons turns out
to have such small impact parameters that the screening of the
electric field around the electron essentially does not prevent
the scatterings. So the cross section for
an electron hitting a phonon
or equivalently a proton the displacement of which represents the phonons
goes as $(\alpha r)^2 \propto \alpha$.
So this cross section is a linear function of  $\alpha$, i.e. it is proportional
to $\alpha$ to the first power (only). This in turn
means that the mean free part $a$ must behave as $\alpha^{-1}$.
The other factors $v$, $c_p$, $\rho$ entering the thermal conductivity
$k=\frac{1}{3}v*a*c_p*\rho$ \cite{Mermin} have no $\alpha$-dependence in our pearl material.
But in ordinary materials the typical electron velocity
at the fermi surface is $\alpha$ (using units in which $\alpha$ is a velocity).

Having argued for how many factors of $\alpha$ must occur in the
expression for the conductivity $k$ also for the pearl matter we can by dimensional analysis
derive the order of magnitude expression for this conductivity.
In fact we simply work as if there was only the velocity $c$
at our disposal and then multiply by the extra dimensionless
factor $\frac{c}{\alpha}$, because we decided that $\alpha $ should
come in to the power $-1$ in $k$. Thus, provided we exclude the possibility that the velocity
of sound can come in, we obtain the result from dimensional analysis:

\begin{eqnarray}
 k &=& \frac{c}{\alpha} * c*p_f^2\nonumber\\
&=& \frac{c^2p_f^2}{\alpha}
\end{eqnarray}

Had we considered ordinary matter at usual pressures, where
the electrons and nuclei remain non-relativistic so that
there is only one relevant velocity $\alpha$, the only
possibility by dimensional analysis
is
\begin{eqnarray}
 k &=& \alpha p_f^2.
\end{eqnarray}

\subsection{More accurate heat conductivity}
To do a little bit better than the above
only dimensional or order of magnitude derivation
of the heat conductivity $k$ for our pearls,
we now consider the supposed main effect of
scattering the electrons - namely scattering by the
emission of phonons - a bit more accurately.
We start from the transition rate as given
in formula (26.40) in the book ``Solid State Physics''
by Ashcroft and Mermin \cite{Mermin}
\begin{eqnarray}
 |g_{\vec{k}\vec{k}'}|^2 &=& \frac{1}{V} \frac{4\pi e^2}{|\vec{k} -\vec{k}'|^2+k_0^2}\frac{1}{2}\hbar
\omega_{\vec{k}-\vec{k}'}\label{fl}\\
&=& \frac{1}{V} \frac{4\pi \alpha}{|\vec{k} -\vec{k}'|^2+k_0^2}\frac{1}{2}\hbar
\omega_{\vec{k}-\vec{k}'}
\end{eqnarray}
 The notation here is
\begin{itemize}
 \item $g_{\vec{k}\vec{k}'}$ is the transition amplitude for an electron going from
having wave number $\vec{k}$ to wave number $\vec{k}'$ by emission (or absorption)
of a phonon of energy $\hbar \omega_{\vec{k}- \vec{k}'}$.
\item The quantization volume $V$ is used to discretize the wave numbers
$\vec{k}$ and $\vec{k}'$.
\item The charge $e$ in the first line is, following Ashcroft and Mermin, in e.s.u. units.
In our notation the potential between two elementary
charges separated by a distance $r$ is $\alpha/r$.
Thus we have $\alpha=e^2$, where $e$ is the
one in the first  line (\ref{fl}).
\item $k_0 =2\sqrt{\frac{\alpha}{\pi c}}p_f$ is the momentum corresponding to the screening length
given by (\ref{f13}).
\end{itemize}

We may consider the transition amplitude $g_{\vec{k}\vec{k}'}$ as a
matrix element of the perturbation energy $V_{int}$, due to the interaction with the
phonon between the electron state $\vec{k}$ and the electron state $\vec{k}'$.
Counting the dimensionality of the $\vec{k} -\vec{k}'$ say as momentum, it is
indeed seen that the quantity $|g_{\vec{k}\vec{k}'}|^2 $ (in the equation (\ref{fl})) has dimensionality
of energy squared.

The Fermi golden rule says that the decay rate $\Gamma_{\vec{k}f}$  of say the
electron in the state $\vec{k}$ decaying into a continuum of
states $f = (\vec{k}', \gamma_f )$ of say a phonon marked $\gamma_f$  and an electron with
wave number $\vec{k}'$ is
\begin{eqnarray}
 \Gamma_{\vec{k} f}&=& 2\pi \rho_{DOS} |g_{\vec{k}\vec{k}'}|^2
\end{eqnarray}
Here $\rho_{DOS} $ is the density of states as a function of energy of the combined
electron phonon states $f$. This factor $\rho_{DOS}$ will come out
effectively, if instead we insert an energy conservation delta-function
$\delta(E_f-E_{\vec{k}})$ and integrate or sum over all the possible
combined states $f$. Thus the total decay rate due to the interaction ending
with the  phonon having the momentum equal to $\vec{k}-\vec{k}'$ is
(ignoring at first the boson statistical
enhancement
by the number of phonons already in the state):
\begin{eqnarray}
 \Gamma_{\vec{k} f}&=& \sum_{\vec{k}'} 2\pi \delta(E_{\vec{k}} - E_f)
|g_{\vec{k}\vec{k}'}|^2\\
&=& \int  2\pi \delta (E_{\vec{k}} - E_f) |g_{\vec{k}\vec{k}'}|^2 \frac{d^3\vec{k}'}{(2\pi)^3}V.
\end{eqnarray}
Here we have used the replacement of the sum over $\vec{k}'$ by an integration
\begin{eqnarray}
 \sum_{\vec{k}'} \rightarrow \int ... \frac{Vd^3\vec{k}'}{(2\pi)^3}.
\end{eqnarray}
If the final state for the electron $\vec{k}'$ is not empty, of course its
contribution will be missing. We are concerned with electrons $\vec{k}$
near the fermi surface where they have the possibility to decay into
empty states $\vec{k}'$. Near the fermi surface we have
$ \frac{dE_{|\vec{k}'|}}{d|\vec{k}'|} = v_{fermi} =c$ in our pearls, and thus
\begin{equation}
 \delta(E_{\vec{k}} - E_{\vec{k}'}) = \frac{1}{v_{fermi}} \delta(|\vec{k}|-|\vec{k}'|).
\end{equation}
The phonon velocity - i.e. the speed of sound (\ref{sound}) - is
of the order $v_{fermi}\sqrt{\frac{m_{e\; rel}}{M_N}}$ as compared to the electron
velocity $v_{fermi}$. Thus the sound velocity is relatively low and the phonon
energy $\hbar \omega_{\vec{k}-\vec{k}'}$ becomes small compared to
the energies of the electrons. Consequently the phonon energy
is not so important in the energy conservation delta-function.

Luckily we do not have to calculate the phonon energy $\hbar \omega_{\vec{k}-\vec{k}'}$
in the high temperature regime we have in mind, because
it is cancelled out by the boson enhancement effect.
This effect enhances the decay rate into a boson state by a factor equal to the number
of bosons already present in that state. It namely happens that, since the
number of bosons/phonons at temperature $T$ in a state with energy $\hbar \omega_{\vec{k}-\vec{k}'}$
is just $T/(\hbar \omega_{\vec{k}-\vec{k}'})$, the phonon energy $\hbar \omega_{\vec{k}-\vec{k}'}$
drops out of the expression for the decay rate:

The decay rate of an electron state with momentum/wave-number $\vec{k}$
under the temperature $T$ and including the boson-enhancement effect
becomes
\begin{eqnarray}
  \Gamma_{\vec{k} f}^{bose-enhanced}
&=& \int  2\pi \delta(E_{\vec{k}} - E_f) |g_{\vec{k}\vec{k}'}|^2
\frac{T}{\hbar \omega_{\vec{k}-\vec{k}'}} \frac{d^3\vec{k}'}{(2\pi)^3}V\\
&=&\int  2\pi \delta(E_{\vec{k}} - E_f) *
 \frac{4\pi \alpha}{|\vec{k} -\vec{k}'|^2+k_0^2}\frac{1}{2}
T \frac{d^3\vec{k}'}{(2\pi)^3}\\
&= & \frac{\alpha T}{2\pi v_{fermi}}\int  \frac{\delta(|\vec{k}| - |\vec{k}'|) }{|\vec{k} -\vec{k}'|^2+k_0^2}
 d^3\vec{k}'\\
&\approx &
\frac{\alpha T}{2\pi v_{fermi}}\int^{|\vec{k}'|=|\vec{k}| \approx p_f}  \frac{1 }{|\vec{k} -\vec{k}'|^2+k_0^2}
 d^2\vec{k}'\\
&=& \frac{\alpha T}{ 2 v_{fermi}}\int_0^{\approx 4p_f^2}  \frac{1 }{|\Delta\vec{k}'|^2+k_0^2}
  d(|\Delta \vec{k}'|^2)\\
&\approx & \frac{\alpha T}{ 2c}\ln \left ( \frac{4p_f^2}{k_0^2}\right ) = \frac{\alpha T}{ 2c}\ln \left ( \frac{\pi c}{\alpha}\right ).
\end{eqnarray}

Here we suppose the velocity of the decaying electron is $v_{fermi} \approx c$.
In the approximation of the just discussed
phonon interaction caused scattering being dominant, the mean free path $a$ becomes
\begin{eqnarray}
 a &=& c/\Gamma_{\vec{k} f}^{bose-enhanced}\\
&\approx&  \frac{2c^2}{\alpha T \ln \left ( \frac{\pi c}{\alpha}\right )}
=  \frac{2c^2}{6.1 \alpha T}.
\end{eqnarray}

Let us now resume the expressions for the quantities $c_p$, $v$, $\rho$ and $a$ to be used to
construct the conductivity $k= \frac{c_p a \rho v}{3}$. We have
\begin{itemize}
 \item The average heat capacity of an electron $c_V \approx c_p= \frac{\pi^2}{2} *\frac{T}{E_f}$.
\item The mean free path - here taken with phonons dominating - $a \approx \frac{2c^2}{\alpha T
\ln \left ( \frac{\pi c}{\alpha}\right )}$
\item The density of electrons in the material $\rho = \frac{p_f^3}{3 \pi^2}$.
\item The electron velocity $\approx c$ in our pearls.
\end{itemize}

Thus we obtain the conductivity estimate for our pearl material:
\begin{eqnarray}
 k &=&  \frac{c_p a \rho v}{3}\label{fk}\\
&\approx&  \frac{   c^2 p_f^2}{9 \alpha  \ln \left ( \frac{\pi c}{\alpha}\right )  }
=  \frac{   c^2 p_f^2}{55 \alpha    }.\label{lk}
\end{eqnarray}

Compared to the expression $k \sim \frac{c^2 p_f^2}{\alpha}$
obtained above just by dimensional and physical arguments,
the present supposedly more accurate expression deviates
from it by a factor $55$ in the
denominator. This factor in the denominator arose from the order unity numerical
numbers included giving 9 multiplied by an extra logarithm
essentially of the inverse $\alpha$.


Let us now compare our calculational method to the non-relativistic physics of
ordinary metals:

In ordinary metals the fermi velocity is no longer $c$ as in the interior
of the dark matter pearls, but rather of the order of $\alpha =\frac{c}{137}$.
So we should just replace the $c$ in our formula (\ref{lk}) for $k$ with true fermi velocity,
approximately $\alpha$. Thus our expectation for the non-relativistic
material thermal conductivity is $k=\frac{\alpha p_f^2}{9ln\pi} = \frac{\alpha p_f^2}{10}$.

Assuming that the fermi momentum is given by a fermi velocity $v_f$ as $m_e v_f$
our dimensional analysis formula means that we predict the ratio $k/v_f^2$ to be
\begin{eqnarray}
\frac{k}{v_f^2} &=&\frac{\alpha p_f^2}{v_f^2} =\alpha m_e^2.
\end{eqnarray}
In rather strange units we have
\begin{eqnarray}
\alpha m_e^2 &=&  1.6*10^{14}m^{-3}s,
\end{eqnarray}
while
\begin{eqnarray}
\frac{W/(m K)}{(10^6 m/s)^2}&=& \frac{1}{1.38}*10^{11}m^{-3}s
\end{eqnarray}
so that
\begin{eqnarray}
\alpha m_e^2 &=& 2250 \frac{W/(m K)}{(10^6 m/s)^2}.
\end{eqnarray}
If we take our best estimate for $k$, which is 10 times smaller than the dimensional estimate for non-relativistic ordinary materials, we get
the prediction
\begin{eqnarray}
\frac{k}{v_f^2} &=& \frac{\alpha m_e^2}{10} =  225 \frac{W/(m K)}{(10^6 m/s)^2}.
\end{eqnarray}
 For a series of metals, copper, gold, silver, iron, lead, lithium, mercury,
 we find respectively the thermal conductivity values $k=$ 385(Cu), 314(Au), 406(Ag), 79.5(Fe) ,34.7(Pb), 85(Li),8.3(Hg), in the
 unit $W/(m K)$ and
 the fermi velocities, 1.57(Cu), 1.40(Au), 1.39(Ag), 1.98(Fe), 1.83(Pb), 1.29(Li), 1.58(Hg), in the unit $10^6 m/s$.
 This means that they have the values,
 156(Cu), 160(Au), 210(Ag), 20(Fe), 13(Pb), 51(Li), 3.3(Hg)
 respectively for the ratio $\frac{k}{v_f^2}$.

 The variation of the ratio $\frac{k}{v_f^2}$, which is a constant in our calculation,
 suggests an uncertainty of this calculation of the order of a factor 5
 up or down. This means our value of $k$ comes with an uncertainty $\exp(\pm 150 \%)$.
 But actually all the mentioned metals except silver have a lower conductivity $k$ than
 our estimate.
 Taking the above metals as representing a mean value for $\frac{k}{v_f^2} \sim 50$,
 we would assume that our a priori estimates should be decreased by a factor $\frac{225}{50} \sim 5$.
 We might expect that our calculational estimate (\ref{lk}) for the conductivity of our pearls
 should be decreased by a similar factor of 5. If so,
the corrected estimate for the conductivity of the relativistic interior of the dark matter would be
 \begin{eqnarray}
 k_{empirically  \:  corrected} &=& \frac{c^2 p_f^2}{250\alpha}. \label{emp}
 \end{eqnarray}
We shall consider this possible correction in section \ref{fits}.

\section{Cooling of Pearl}

Let us form ourselves a picture of how we imagine the
cooling goes when two pearls have collided:

\begin{itemize}
 \item First notice that the skin of the
two pearls having collided will typically deliver the energy/heat in a
highly non-uniform way. It is basically that some very thin region near the
part of the skin that gets most contracted takes up almost all the
heat $\sim E_S$. In the very first moment before heat has had time to spread
therefore of course the rest of the pearl is cold.
\item Next the heat spreads according to heat diffusion. The easiest situation for calculation
might be the special case where only one point was heated up, while a more realistic picture may be that it is
a smaller part of the skin that gets heated. If a unit amount of heat were
sitting at the origin,
with delta-function density $u = \delta(\vec{r})$, say
at time $t=0$,
it is well-known and easy to see that this situation corresponds to the solution of the heat
diffusion equation
   \begin{eqnarray}
\rho c_p \dot{u} - k \Delta u &=&0,
\end{eqnarray}
where we have no extra addition  of heat and
$\dot{u}= \frac{\partial u}{\partial t}$
is the time derivative of the energy
density $u$, while $\Delta$ is the
Laplacian.
The solution of the heat diffusion equation with
the initial condition (at
time $t=0$)
\begin{equation}
u= \delta (\vec{r})
\end{equation}
leads to the heat-kernel:
\begin{equation}
u=\left ( \frac{4\pi kt}{\rho c_p}
\right )^{-3/2}\exp\left(-\frac{\vec{r}^2}{\frac{
4kt}{\rho c_p}}\right).\label{kerneln}
\end{equation}
From this solution one sees that the size of the
region, which gets hot, spreads - defining its border say
from where the expression in the exponent is just unity -
with
\begin{eqnarray}
 |\vec{r}_{border}| &\propto& \sqrt{t}.
\end{eqnarray}
That is to say the size $|\vec{r}_{border}|$ of this hot
region grows at first very fast, and then slows down.
\item The time it takes for the ``hot'' blob  of temperature T =
81 MeV
or T =
162 MeV to reach the size of the pearl, i.e. to
reach $|\vec{r}_{border}| \sim R = 0.05$ cm
or $ \sim R = 0.025$ cm will be of the order
\begin{eqnarray}
 t_{spread} &\approx& \frac{\rho c_p}{4k}*R^2\label{tsp}\\
&\approx& \frac{\frac{ p_f^3}{3\pi^2}* \frac{T\pi^2}{2E_f}}{4 *\frac{c^2p_f^2}{55\alpha}}*R^2  
\, = \, \frac{\alpha 55 R^2 T}{24c^3}\\
&\approx &\frac{55(
0.05\ cm)^2(
81\ MeV)}{137*24c^2}\quad \hbox{($\phi_H$=123\, GeV)}\\
&=&
 5.2*10^{-3}\, s \qquad \hbox{($\phi_H$=123 GeV)}
\end{eqnarray}
or
\begin{eqnarray}
t_{spread}
&=&
2.6*10^{-3}\,s
\qquad \hbox{($\phi_H$=0)}
\end{eqnarray}

\item During a time of the order of
$t_{spread}$ there will be a hot spot spreading.
Since this spot has a Gaussian temperature profile
(\ref{kerneln}), the temperature in the outskirts of it
as a function of the distance will quickly drop down to be almost as cold as the
initial cold pearl, which had been in balance with the
3 K outer space for milliards of years. We shall thus
imagine an intermediate situation in which part of
the pearl is still exceedingly cold, while there is a hot spot
with temperature close to the average temperature
of the pearl of the order
81 MeV or
162 MeV.
\item Where this hot spot reaches the surface of the pearl,
there will be radiation into space with frequencies
up to the average temperature of order, say,
81 MeV or 162 MeV.
That will however then quickly lower the temperature
just near the surface and the hot spot will be effectively
pushed a bit closer to the center of the pearl.
\item During this period of there being a rather isolated hot spot
there will from the border of this hot spot, where the temperature is
just of the order
of 3.5 keV, still exist to the cold side an undisturbed homolumo gap.
Thus only radiation of frequencies lower than the homolumo
gap size can penetrate through the cold part of the pearl
out into outer space. Just where the temperature passes order of magnitude-wise
the 3.5 keV, there will be some excited electrons in the material
able to interact so strongly with the photons with frequencies below the
gap value
that there will be emitted radiation in these frequencies with
approximately the strength of a black body.

Provided the surface specified by having just the
suitable temperature close to 3.5 keV, from which the
3.5 keV photons just could  reach optically from the outside, dominates
the surface around the hot spot, there will be a
dominantly 3.5 keV radiation cooling off of the hot spot in such
an era wherein part of the pearl is still
very cold.

In order for such a dominating 3.5 keV emission, it is
needed that there is no hot edge of the hot spot
reaching the surface of the pearl. At such an edge there would
namely be radiated extremely much energy because of the $T^4$
proportionality of the emission.  But one could hope that
such a hot edge region would be cooled off and the hot spot
go into a state with colder and thus 3.5 keV emitting surface
almost all around it.

But even if the temperature at the pearl surface does not
get down to the 3.5 keV order completely, it would be enough
to get a major part of the energy out as 3.5 keV radiation
provided the hot surface pieces are rather small and
the temperature not much bigger than the 3.5 keV.

\item As time passes there is now the possibility that
the hot spot spreads all over the pearl and at the end there will
no longer be any surface where the temperature is just around 3.5 keV.
In such a case the radiation in this frequency will stop or rather
there will no longer be any peak at this frequency in the radiation from
pearl. This kind of stopping happens roughly after the time
$t_{spread}$.

\item But there is another time parameter which we call
$t_{radiation}$ which we define to be the time it
takes to radiate out - say into 3.5 keV - essentially
all the energy in the heat. In the case that this radiation time
should happen to be smaller than the $t_{spread}$,
the energy would run out before the hot spot has
spread all over the pearl. Thus it could be that the
3.5 keV radiation would never be stopped before all the
energy was essentially used up.

\item When finally the energy of the heat from the collision
is about to run out, the hot spot will of course be colder
and there will (again) appear a surface around it where
the temperature is down to about the 3.5 keV and
3.5 keV radiation in a peak will be re-established for
a time.

\end{itemize}

To get an idea as to whether there is any chance
that a scenario like the above could provide enough
3.5 keV radiation so as to
dominate the radiation of the energy from the collision of two pearls,
we shall now estimate the time scale
$t_{radiation}$ for the energy to be used up by this radiation.

Remembering that the energy deposited as heat by the
skin contraction from a collision of two pearls was estimated to be (\ref{ES1})
\begin{eqnarray}
 E_S &=& \frac{3}{16}\frac{Mp_f}{m_N}
\end{eqnarray}
it follows that
\begin{eqnarray}
E_S & = &
5.4 \% M c^2 \qquad \hbox{(with $\phi_H=123\, \hbox{GeV}$)}\label{ES}\\
E_S &=&
10.8 \% M c^2 \qquad \hbox{(with $\phi_H=0$)}
 \end{eqnarray}
Now the mass of the pearl is
\begin{eqnarray}
Mc^2
&= &1.4 *10^8\, \hbox{kg}\, = \,
0.8 *10^{41}\, \hbox{keV}.
\end{eqnarray}
So we get the deposited energy to be
\begin{eqnarray}
 E_S &=&
4*10^{39}\, \hbox{keV} \qquad \hbox{(with $\phi_H=123\, \hbox{GeV}$)}\\
 E_S &=&
8*10^{39}\, \hbox{keV} \qquad \hbox{(with $\phi_H=0 $)}.
\end{eqnarray}
The typical area through which this heat energy
hopefully gets radiated out with a temperature
close to the 3.5 keV - if we shall be successful to have
most of the energy expended in this way - is
the surface area of the pearl
\begin{eqnarray}
 4\pi R^2 &=& 4\pi (
0.05\ \hbox{cm})^2 \qquad \hbox{(for $\phi_H=123\, \hbox{GeV}$)}\\
&=&
7.7*10^{13}\ \hbox{(keV/c})^{-2} \qquad \hbox{(for $\phi_H=123\, \hbox{GeV}$)}
\end{eqnarray}
while
\begin{eqnarray}
 4\pi R^2
&=&
2*10^{13}\ \hbox{(keV/c)}^{-2} \qquad \hbox{(for $\phi_H=0 $)}.
\end{eqnarray}
Taking it that the temperature at the emission layer
is just 3.5 keV and the Stefan constant
$\sigma_{St} = \frac{\pi^2}{60 c^2}$ the radiation of the energy $E_S$
through the area $4\pi R^2$ takes the time
\begin{eqnarray}
 t_{radiation}&=& \frac{E_S}{4\pi R^2 \sigma_{St} T^4}\\
 &=& \frac{
4*10^{39}keV}{
7.7*10^{13}(\hbox{keV/c})^{-2} *\frac{\pi^2}{60 c^2} *(3.5\ \hbox{keV})^4}\\
&=&
1.4*10^6 \, s \qquad \hbox{(for $\phi_H=123$ GeV)}.
\end{eqnarray}
while
\begin{eqnarray}
 t_{radiation}
&=&
1.1*10^7 \, s \qquad \hbox{(for $\phi_H=0 $)}.
\end{eqnarray}
As a comment on the uncertainty of the here evaluated $t_{radiation}$ we especially
worry about the factor
of (3.5 keV)$^4$. This is the fourth power of
the supposed temperature at the place from which the 3.5 keV radiation
is sent out and it is, of course, sensible to take this
temperature at emission of the line to be close to 3.5 keV. But
really the line is supposed to be emitted by exciton decay (see section \ref{why})
and that emission could in principle be done at e.g. a higher temperature.
If in some layer the temperature $T$ is higher, the emission will
be favoured by a factor $T^4$. Thus the effective or dominant temperature is expected to
be bigger than $3.5$ keV. We have only very crude speculative estimates for
how much bigger this could be. However let us say that the truly relevant emission temperature to be used
in evaluating $t_{radiation}$ is
1.5 *3.5 keV rather than just 3.5 keV:
\begin{eqnarray}
 t_{radiation}&\propto& \frac{1}{(3.5\ \hbox{keV})^4}\quad \rightarrow
\quad t_{radiation}\propto \frac{1}{(1.5 *3.5\ \hbox{keV})^4}\label{1c5}
\end{eqnarray}

 This temperature is rather uncertain since
it will depend on how quickly as a function of temperature the homolumo gap
gets washed out. We therefore assign  an uncertainty of at least $\exp(\pm 100\%)$ to
the value of the temperature of emission.
Because this emission temperature comes in to the fourth
power, it follows that $t_{radiation}$ itself has a rather large
uncertainty of $\exp(\pm 400\%)$.

After this discussion we write the final values
\begin{eqnarray}
 t_{radiation} &=&
1.4*10^6s*1.5^{-4}\exp(\pm 400\%)\\
&= &
2.8 *10^5s \exp(\pm 400\%) s \quad \hbox{(for } \phi_H =123\ \hbox{GeV} ).
\label{trad123}\\
 t_{radiation}
&=&
2.2 *10^6\exp(\pm 400\%) s \quad \hbox{(for } \phi_H =0 )\label{trad0}
\end{eqnarray}
In the light of great arbitrariness of our guessed correction by a factor
1.5 in the emission temperature, we do not think it is reasonable to make the
following calculations depend on it. We shall thus go on with the
more well-defined values where we use precisely 3.5 keV for the emission temperature,
(although we do believe that inclusion of the 1.5 factor is more likely to
be right).
In the light of the $\exp(\pm 400\%)$ uncertainty however this makes little difference to our results.

So the time to spread the heat over the whole pearl relative
to the time for emitting the whole energy (if it were done
by 3.5 keV radiation with a temperature of that order too) using the
critical size ball parameters
becomes
\begin{eqnarray}
\frac{t_{spread}}{t_{radiation}}&=&
\frac{
5.2*10^{-3}}{
1.4*10^6s}\\
&=&
\frac{1}{2.7*10^8}*\exp(\pm 400 \%) \qquad  \hbox{(for $\phi_H=123\, \hbox{GeV}$)}. \label{tratio1}
\end{eqnarray}
while
\begin{eqnarray}
\frac{t_{spread}}{t_{radiation}}&=& \frac{
2.6*10^{-3}s
}{
1.1*10^{7}s}\\
&=& \frac{1}{
4.2 *10^{9}}*\exp(\pm 400\%) \qquad \hbox{(for $\phi_H=0 $)}. \label{tratio2}
\end{eqnarray}

After the time $t_{spread}$ the hot spot heated up by the
collision
will have spread to the whole
pearl and no regions with the original low temperature would be left,
Thus after this time $t_{spread} $ the emission frequencies would
be appreciably larger than 3.5 keV and from that time on
 no more 3.5 keV peak radiation can be radiated. This means that
the energy fraction emitted in the line 3.5 keV can at the most
be of the order of $t_{spread}/t_{radiation}$ times the
total emitted energy. If the total energy $E_S$ was emitted only
as 3.5 keV radiation, the number of photons emitted would be given by
\begin{eqnarray}
 \frac{E_S}{3.5\ \hbox{keV}} &=& N_{all\rightarrow 3.5}.
\end{eqnarray}
So the actual number $N$ of 3.5 keV photons emitted could at most be
\begin{eqnarray}
 N&=& \frac{t_{spread}}{t_{radiation}}* \frac{E_S}{3.5\ \hbox{keV}}.
\end{eqnarray}
This in turn would reduce the intensity or say $\frac{N\sigma}{M^2}$,
by being multiplied
by the factor $\frac{t_{spread}}{t_{radiation}}$ =
$\frac{1}{2.7*10^8}$ (for $\phi_H=123\, \hbox{GeV}$) and
$\frac{1}{4.2*10^9}$ (for $\phi_H=0$).
That would make our prediction (\ref{prediction}) even smaller
by a large factor of
$2.7*10^8$ or $4.2*10^9$.
Therefore our fit, with the parameters corresponding to a pearl of
the critical size just needed for stability against collapse, is not so good.

In section \ref{size} we shall give up this critical size assumption
and allow the actual radius of the pearl to be bigger than the critical
radius by some factor $\xi$.
It will turn out that the most important quantities approximately
only depend on the ratio $\xi/\Delta V$. Thus we shall fit
our model with this ratio and a separate value for $\Delta V$ will not be so very
important in the fitting.


\subsection{Why a line? \label{why}}
In the picture just sketched we at least see that as long as
only part of the pearl has been heated up there is a cold part, through
which the X-ray radiation from the hot spot has to penetrate to reach
the outer space and thereby us. Through this cold part of the pearl there
will for radiation with frequencies less than the homolumo gap $E_H$ of eq. (\ref{EHW})
 be essentially free passage as if it were glass, while for
radiation of frequency bigger than $E_H$
the pearl will be highly non-transparent. If for instance there were at
the hot spot surface produced radiation with a normal black body
radiation with the frequency distribution of the Planck radiation
being proportional to $\nu^3$ (where $\nu$ is the frequency), then after
the passage of the cold region of the pearl the spectrum would look like such
a $\nu^3$ spectrum chopped off at the homolumo gap frequency $E_H$. Since $3$
is a rather high power, this would actually mean a spectrum with a
strong maximum approximately at $E_H$
already not completely far away from what is seemingly seen by the
X-ray spectrometers. However, the peak would not be a very sharp
peak in as far as it would have fall off to the left only as a third
power, which is somewhat sharp but not extremely so.

However, we have already mentioned that there is the possibility of obtaining
X-ray radiation at the frequency of the homolumo gap value $E_H$, or almost so,
by excitons - bound states of a hole and an (excited quasi) electron -
decaying into a photon and a phonon. Now the question of course is
whether a large part of the heat energy being converted to radiation
will be produced via such excitons or at least by hole electron
annihilation.

In order to investigate this question as to how big a fraction
of the heat energy comes out from exciton or hole electron annihilation
relative to the part coming out just from thermal radiation we shall first look
at in what ratio the heat energy is transported along in the pearl-material
in the presence of a temperature gradient $\nabla T$ by ordinary
heat conduction
\begin{eqnarray}
\vec{J}_{heat} &=& k \nabla T
\end{eqnarray}
or by means of the holes and electrons moving along in the
material, which also gives effectively an energy current
\begin{eqnarray}
\vec{J}_{exc}&=&\frac{E_H}{-e} *\frac{1}{2}
(\vec{J}_{el} +\vec{J}_{holes}).\label{flow}
\end{eqnarray}
Here $\vec{J}_{el}$ and $\vec{J}_{holes}$ are the currents of electron excitations
and of holes respectively. Since each pair of an electron and a hole brings
along an energy equal to the homolumo gap $E_H$
the current called $\vec{J}_{exc}$ is indeed the current of energy
carried along in the form of electron hole excitations.
Both currents  $\vec{J}_{el}$ and $\vec{J}_{holes}$ are basically Seebeck
or thermoelectric currents and as such proportional to the temperature
gradient $\nabla T$ and thus also proportional w.r.t.
direction and strength of flow to heat current $\vec{J}_{heat}$ although
with a temperature dependent coefficient.
The denominator $-e$ were inserted because the usual Seebeck coefficients
are normalized to give the current including the charge factor $-e$ for an
electron, but we have to count simply the number of electrons, if we want
to get the energy of annihilation be just $E_H$.

\subsubsection{Mott Formulas}

The current, which we are interested in is not exactly the electric
Seebeck current because we want to {\em add} up the flow of holes and
electron excitation, whereas the true electric current is rather the
difference between the two because holes and electrons have
opposite electric charge. So if we use the Seebeck coefficients for
contribution from the bottom of the conduction band $S_C$ and
from the top of the valence band $S_V$ as separate Seebeck
coefficients we should combine them with an opposite sign relative
to the one used when one wants to make the electric full Seebeck
coefficient. Indeed while the usual Seebeck coefficient is:
\begin{equation}
S=\frac{\sigma_C S_C + \sigma_V S_V}{\sigma_C + \sigma_V}
\end{equation}
the coefficient important for us is:
\begin{equation}
S_{our} = \frac{\sigma_C S_C - \sigma_V S_V}{\sigma_C + \sigma_V}.
\end{equation}
Indeed the flow of energy in the form of excitations - electrons and
holes - formula (\ref{flow}) becomes
\begin{eqnarray}
\vec{J}_{exc}&=&
\frac{E_H}{-e}*\frac{1}{2}(\sigma_C+\sigma_V) S_{our} \nabla T\\
&=&
\frac{E_H}{-e}*\frac{1}{2}(\sigma_C S_C -\sigma_V S_V) \nabla T,
\end{eqnarray}
where $S_C$ and $S_V$ are the Seebeck coefficients as if there were
respectively only the conduction band or only the valence band.

Now we insert the Mott-formulas for these Seebeck coefficients and
conductivities $\sigma_C$ and $\sigma_V$ as found in the Wikipedia
article on ``Seebeck coefficients'',
\begin{eqnarray}
S_C &=& \frac{k_B}{-e}\left [\frac{E_C-\mu}{k_BT} +a_C +1\right ]\\
\sigma_C &=& A_C (k_BT)^{a_C}\exp(-\frac{E_C-\mu}{k_BT}) \Gamma(a_C+1)\\
S_V &=& \frac{k_B}{e}\left [\frac{\mu-E_V}{k_BT} +a_V +1\right ]\\
\sigma_V &=& A_V (k_BT)^{a_V}\exp(-\frac{\mu-E_V}{k_BT}) \Gamma(a_V+1).
\end{eqnarray}
Here $E_C$ is the lowest energy point of the conduction band and
$E_V$ the highest point of the valence band, so that the homolumo
gap is
\begin{equation}
E_H = E_C -E_V,
\end{equation}
and $\mu$ is the chemical potential for electrons.
The conductivity band levels' conductivity function $c_C(E)$
has been put to the approximate form
\begin{equation}
c_C(E) = A_C (E-E_C)^{a_C}
\end{equation}
and analogously for the valence band
\begin{equation}
c_V(E) = A_V (E_V-E)^{a_V}.
\end{equation}
The parameters $A_C,A_V,a_C, \hbox{ and } a_V$ are material dependent
ansatz parameters with the small $a$'s in the range 1 to 3. The
conductivity functions may be defined as
\begin{eqnarray}
c(E) &=& e^2 D(E) \nu(E),
\end{eqnarray}
where $D(E)$ is diffusion constant for electrons of energy $E$
and $\nu (E)$ the level density at $E$ This $c(E)$ represents the density
on the energy axis of conductivity.

For our estimation we shall like to make a symmetric approximation
because we believe that the homolumo gap $E_H$ is so narrow compared to the
fermi energy $E_f$ that very little asymmetry will appear. That is to say
we take $a_C = a_V =a$ and $A_C = A_V =A$, and in addition the chemical
potential $\mu$ in the middle of the homolumo gap.
Then we get
\begin{eqnarray}
S_{our} &=&  \frac{\sigma_C S_C - \sigma_V S_V}{\sigma_C + \sigma_V}\nonumber \\
&=& \frac{1}{2} ( S_C -S_V)\\
&=&  \frac{k_B}{-e}\left [
\frac{E_H}{2k_BT} +a +1\right ].
\end{eqnarray}
and now we shall use the Wiedemann Franz law relating the conductivity
$\sigma = \sigma_C + \sigma_V$ to the heat conductivity $k$ by
\begin{equation}
\frac{k}{\sigma} = L T
\end{equation}
where
\begin{equation}
L = \frac{\pi^2}{3}*\left ( \frac{k_B}{e}\right ) ^2.
\end{equation}

The point now is that since we have two types of energy flow
both flowing proportionally to the temperature gradient $\nabla T$,
we can consider the ratio of the coefficients to these gradients.
We have
\begin{eqnarray}
\vec{J}_{heat} &=& k \nabla T\\
\vec{J}_{exc} &=& \frac{1}{-e} \sigma*
E_H*S_{our} \nabla T\\
& = & \frac{3k}{\pi^2}\left [  \frac{1}{2}\left (
\frac{E_H}{k_BT}\right )^2 +(a +1)
\frac{E_H}{k_BT}\right ]\nabla T.
\end{eqnarray}

This means that the ratio of the part of energy carried via the excitation electrons and the holes
relative to the normal heat conduction heat flow is

\begin{eqnarray}
\frac{\hbox{``electron and hole carried''}}{\hbox{``normal heat flow''}} &=&
\frac{3}{\pi^2}\left [\frac{1}{2}\left(
\frac{E_H}{k_BT}\right )^2
+(a +1)
\frac{E_H}{k_BT}\right ].\nonumber
\end{eqnarray}

We see that for high temperature compared to the homolumo gap the
normal heat conduction dominates, but that when temperature goes small,
the energy transport by means of excited electrons and holes which
finally get radiated out as 3.5 keV radiation takes over.

Now the situation during the time when the pearl is still cold around  the
hot spot is such that only  electromagnetic radiation with
lower or equal frequency with the homolumo gap can escape to the outside.

But if it comes as it must from the neighboring region with temperature
approaching the homolumo gap the mechanism with the annihilating
excited electrons and holes take over.

So it seems that indeed during the spreading time $t_{spread}$ of the
hot spot during which it grows up to cover the whole pearl,
the 3.5 keV radiation peak must be the dominant emission
process. So if just this grow up
of the hot spot time $t_{spread}$ were long enough for the energy
to be emitted the main energy would be emitted into the line 3.5 keV.

However, with the parameters of a critical size pearl, it seems that the time to emit the
energy $t_{radiation}$ is longer than the spreading time $t_{spread}$ by a
factor of
$2.7*10^8$  or $4.2 *10^9$.


\section{Dropping critical size of pearls. \label{size}}

Now we must remember that in our previous article \cite{Tunguska},
we made the assumption that the size of the pearls was just so that
they were on the borderline of collapsing by squeezing out the
contained nuclei or nucleons. Of course it would be highly unlikely
that they should be exactly on this border. They would have to be at
least a little bit
bigger than that. Instead of really making this assumption, we shall now
use the experimentally determined quantity $\left(\frac{N\sigma}{M^2}\right)_{exp}$
given in eq. (\ref{Freyf}) to
fit the ratio
\begin{eqnarray}
 \xi &=& \frac{R_{actual}}{R_{crit}}
\end{eqnarray}
of the actual average radius of the pearls $R_{actual}$ compared to
the radius $R_{crit}$ that corresponds to the pearls being in the
critical state just about to collapse. This borderline to collapse
corresponds to taking the Fermi momentum
\begin{equation}
 p_f = 2\Delta V
 \end{equation}
now written as:
\begin{equation}
p_{f \; crit}= 2\Delta V.
\end{equation}

Now we have shown earlier (\ref{R}) that the radius $R$ and the Fermi momentum $p_f$
of a pearl satisfy the relation
\begin{eqnarray}
Rp_f &=& \left(
\frac{9\pi M}{8m_N}\right)^{1/3}. \label{Rpf}
\end{eqnarray}
So if we keep the mass $M$ of the pearl fixed, which is essentially given by the rate of
Tunguska events, the actual Fermi momentum becomes
\begin{equation}
p_{f \; actual} = \xi^{-1} p_{f \; crit}.
\end{equation}
In addition we can and do keep
$\Delta V$ fixed under the $\xi$ modification.

Inserting the above relation (\ref{Rpf}) into the pressure balance equation (\ref{pressure})
between the surface tension of the pearl and the relativistic electrons inside we find
\begin{equation}
S = p_f^3\frac{1}{24\pi^2}\left(\frac{9\pi M}{8m_N}\right)^{1/3}.
\end{equation}
So the tension for a critical size pearl of mass $M = 1.4*10^8$ kg \cite{Tunguska}
with $p_{f \, crit} = 2\Delta V = 270$ MeV is given by
\begin{equation}
S^{1/3}_{crit} = p_{f \, crit}\left(\frac{1}{24\pi^2}\right)^{1/3}\left(\frac{9\pi M}{8m_N}\right)^{1/9}
= 380 \, \hbox{GeV}.
\end{equation}
It follows that
\begin{eqnarray}
 S_{actual}^{1/3} &=& S_{crit}^{1/3}/\xi. \label{Sactual}
\end{eqnarray}

It is remarkable that the parameter $\frac{N\sigma}{M^2}$ describing
the overall scale of the intensity of the 3.5 keV line radiation
is very sensitive
to this here introduced parameter $\xi$. In fact it is proportional to $\xi^6$ as we show below,
provided the time ratio  $\frac{t_{spread}}{t_{radiation}} \le 1$ (if not it is simply proportional to $\xi$):

The $\Delta V$ and $\xi$ dependence of some of our quantities
 for a fixed pearl mass $M$ are as follows :
\begin{eqnarray}
\hbox{Pearl radius } R &\propto& \frac{\xi}{\Delta V} \\
\hbox{Cubic root of tension } S^{\frac{1}{3}} &\propto& \frac{\Delta V}{\xi} \\
\hbox{Fermi momentum } p_f &\propto& \frac{\Delta V}{\xi}\\
\hbox{Energy release by collision } E_S &\propto&  \frac{\Delta V}{\xi}\\
\hbox{Collision cross section } \sigma &\propto& \left ( \frac{\xi}{\Delta V }\right )^2\\
t_{spread} &\propto& \frac{\xi^2}{\Delta V}\\
t_{radiation} &\propto& \left (\frac{\Delta V}{\xi} \right )^3\\
\frac{t_{spread}}{t_{radiation}} &\propto& \frac{\xi^5}{\Delta V^4}\\
\left.\frac{N\sigma}{M^2}\right |_{all E_S \rightarrow 3.5 keV} &\propto& \frac{\xi}{\Delta V}\\
 \frac{t_{spread}}{t_{radiation}} *\left.\frac{N\sigma}{M^2}\right |_{all E_S \rightarrow 3.5 keV} &\propto& \frac{\xi^6}{\Delta V^5}
\end{eqnarray}

In fact our true prediction
for this cross section for pearl collision $\sigma$  times the number $N$ of 3.5 keV
photons per collision divided by the pearl mass squared $M^2$ is given by
\begin{eqnarray}
 \left. \frac{N\sigma}{M^2}\right |_{pred} &=& \min\left \{1,\frac{t_{spread}}{t_{radiation}}\right \}*\left.\frac{N\sigma}{M^2}\right
|_{\hbox{as if $E_S \rightarrow 3.5 keV's$}}
\end{eqnarray}
 Here for a critical size ball we have (\ref{ES1})
\begin{eqnarray}
 E_S &=& \frac{Mp_{f \; crit}}{m_N}*\frac{3}{16},
\end{eqnarray}
and we define
 $\left.\frac{N\sigma}{M^2}\right |_{\hbox{as if } E_S\rightarrow 3.5 keV's}$
as the value for the ratio $\frac{N\sigma}{M^2}$ calculated as if all the energy
$E_S$ from the surface contraction went into 3.5 keV photons.
The factor $\min \left \{1, \frac{t_{spread}}{t_{radiation}} \right \}$ gives the ratio of
the energy emitted as 3.5 keV radiation compared to the total energy released
$E_S$. For $t_{spread} \ge t_{radiation}$ all the energy gets radiated
from a surface cold enough that it all becomes 3.5 keV radiation.

Now we introduce the $\xi$ parameter to allow
for the radius $R_{actual}$ not being the critical one
just corresponding to collapse limit and obtain
\begin{eqnarray}
 \left. \frac{N\sigma}{M^2}\right |_{actual} &=&
\min \left  \{1, \left (\frac{t_{spread}}{t_{radiation}}\right )_{actual}\right \}*
\left. \frac{N\sigma}{M^2}\right |_{{actual, \, as \, if\, E_S\rightarrow 3.5 \,keV's}}\nonumber\\
 &=& \min \left \{1,\left (\frac{t_{spread}}{t_{radiation}}\right )_{crit}\xi^5 \right \}\; * \;
\left .\frac{N\sigma}{M^2}\right |_{{crit,\,  as \, if \, E_S \rightarrow 3.5\, keV's}}\xi \nonumber\\
&=& \left (\frac{t_{spread}}{t_{radiation}}\right )_{crit} *
\left .\frac{N\sigma}{M^2}\right |_{{crit, \, as \, if\, E_S \rightarrow 3.5\, keV's}}\xi^6 \nonumber\\
&&\hbox{(provided $\frac{t_{spread}}{t_{radiation}}\le 1$)}\nonumber\\
&=& \frac{1}{2.7*10^8}*2.0*10^{21}\xi^6 \ \hbox{cm$^2$/kg$^2$}
\ \hbox{(for $\phi_H=123$ GeV)} \label{actual1}
\end{eqnarray}
or
\begin{eqnarray}
\left . \frac{N\sigma}{M^2}\right |_{actual}&=&\min\left \{1, \frac{\xi^5}{
4.2*10^9}\right \} *
1.0 *10^{21}\xi\ \hbox{cm$^2$/kg$^2$}  \ \hbox{(for $\phi_H=0$)} \label{actual2}
\end{eqnarray}
Here we have used (\ref{prediction}), (\ref{tratio1}) and (\ref{tratio2}) for the critical size ball quantities.
These predictions (\ref{actual1}) and (\ref{actual2}) should be compared with the experimental value (\ref{Freyf}):
\begin{eqnarray}
 \left.  \frac{N\sigma}{M^2} \right |_{exp} &=& (1.0 \pm 0.2) *10^{23}\ \hbox{cm$^2$/kg$^2$}.
\end{eqnarray}
So requiring $\left . \frac{N\sigma}{M^2}\right |_{actual} = \left.  \frac{N\sigma}{M^2} \right |_{exp}$ gives
\begin{eqnarray}
\xi &=& \sqrt[6]{1.35 *10^{10}}=49
\quad \hbox{(for $\phi_H=123$ GeV)} \label{xi123}
\end{eqnarray}
or
\begin{eqnarray}
\xi &=& \sqrt[6]{4.2 *10^{11}}=87
\quad \hbox{(for $\phi_H=0 $)} \label{xi0}
\end{eqnarray}

We note that both of  these $\xi$-values
 leave the ratio
$\left. \frac{t_{spread}}{t_{radiation}}\right |_{actual}\approx 1$ so that the simple
$\xi^6$-corrections (\ref{actual1}, \ref{actual2}) would still be o.k.

With the corrections (\ref{xxi123}) and (\ref{xxi0}) of section \ref{further} these $\xi$-values become
\begin{eqnarray}
  \xi &=& 1.3*
49 = 64
\quad \hbox{(for $\phi_H=123$ GeV)}\label{xifit123}
\end{eqnarray}
or
\begin{eqnarray}
\xi &=& 1.4*
87 =120
\quad \hbox{(for $\phi_H=0 $)}.\label{xifit0}
\end{eqnarray}
The energy gap is now predicted to be reduced from its critical value $E_{H \; crit}$
(\ref{EHold1}) or (\ref{EHold2}) to
\begin{eqnarray}
  E_{H} &=& E_{H \; crit} /\xi \nonumber \\
&=& \frac{240 \; keV}{49} = 4.8 \exp(\pm 100\%)\  \hbox{keV} \quad \hbox{(for $\phi_H=123$ GeV)}
\end{eqnarray}
or
\begin{eqnarray}
E_H &=&
\frac{480 \; keV}{87}= 5.6 \exp(\pm 100\%)\  \hbox{keV} \quad \hbox{(for $\phi_H=0 $)}.
\end{eqnarray}
Here we have introduced a logarithmic error of 100\% to reflect the expected order of
magnitude accuracy of our calculation of the energy gap based on the Thomas Fermi model.

The corresponding correction (\ref{Sactual}) to the surface tension $S$
then gives
$S^{1/3} = 380\, \hbox{GeV}/49 =
8 \, \hbox{GeV}$ for $\phi_H=123\, \hbox{GeV}$ and
$S^{1/3} = 760\, \hbox{GeV}/87 = 9$ GeV for $\phi_H=0$.


In the appendix \ref{RAA} we shall investigate some theoretical expectations for this
tension $S$ from the effective potential as a function of the relevant scalar fields, the Higgs field
$\phi_H$ and the effective field $\phi_F$ for the bound state supposed to be the
main constituent in the ``condensate'' phase. The a priori value from this theory
turns out to be $S^{1/3}_{theory} \approx 100$ GeV, which is embarrassingly much
larger than these fitted values. However, we shall put forward the idea that
there is yet another vacuum phase, which enforces the effective potential
to be flatter than we assumed when obtaining the above estimate $S^{\frac{1}{3}}_{theory} \approx 100$ GeV.
In this way we get what we call ``theory 2)'' for the tension $S$, for which our estimate becomes
$S^{1/3} \approx 30$ GeV. Even this value is large compared to the above fitted values
around
8 GeV.
However in the following section we investigate a more detailed fitting procedure
using the
only relevant combination of parameters $\frac{\xi*10 MeV}{\Delta V}$.

\section{Fitting with $\left. \frac{t_{spread}}{t_{radiation}}\right |_{actual}\ge 1$.\label{fits} }

We remarked above that actually the ratio $\frac{t_{spread}}{t_{radiation}}$, giving
the fraction of the energy from the contraction of the surface ending up in the
3.5 keV radiation, turned out to be very close to unity in our fitting.
However the calculation of $t_{radiation}$ is very sensitive to the emission
temperature and we have assigned an uncertainty of $\exp(\pm 100\%)$ to the
emission temperature. So the uncertainty on $t_{radiation}$ becomes $\exp(\pm 400\%)$.
Also the calculation of $t_{spread}$ depends on the hard to calculate heat conductivity.
The time ratio $\frac{t_{spread}}{t_{radiation}}$ is therefore very
uncertain (by $\exp(\pm 400\%)$). However, if this
ratio happens to be bigger than unity the expression $\min \left \{1, \frac{t_{spread}}{t_{radiation}} \right \}$
simply becomes unity. In this case the uncertainty from the
calculation of the ratio disappears from the calculation and we get a much
slower variation with the parameter $\xi$ of the predicted value of $\frac{N\sigma}{M^2}$,
which is to be compared to (\ref{Freyf}).
Thus, in this case, there is a much bigger range of $\xi$ available for fitting than
in the range of $\xi$ where $\frac{t_{spread}}{t_{radiation}}$ ratio is smaller than unity.
From the point of view that the true value of $\xi$ can be considered random, there
is therefore an appreciably higher chance for $\xi$ to lie in the range where
$\frac{t_{spread}}{t_{radiation}} \ge 1$

According to (\ref{tratio1},\ref{tratio2}),
under the assumption of exactly critical size pearls (i.e. for $\xi= 1$),
 the ratio $\frac{t_{spread}}{t_{radiation}}$ takes the values
$\frac{1}{2.7*10^8}$ and $\frac{1}{4.2*10^9}$ for $\phi_H =123$ GeV and $\phi_H=0$ respectively.
This means that this ratio becomes just unity for
\begin{eqnarray}
 \xi &=& \xi_1 = \sqrt[5]{2.7*10^8} = 49 \quad \hbox{(for $\phi_H=123\, \hbox{GeV}$)} \label{xi1crit1}\\
 \xi &=& \xi_1 = \sqrt[5]{4.2*10^9} = 84 \quad \hbox{(for $\phi_H=0$)} \label{xi1crit2}.
\end{eqnarray}
Once we look for $\xi$-values larger than $\xi_1$,
the sensitivity of the observable quantity $\frac{N\sigma}{M^2}$ to $\xi$ becomes much weaker (only
depending on $\xi$ to the first power). In practice we could for the matter of fitting the
3.5 keV rate simply take $\xi=\xi_1$.

Statistically it is very possible that our estimates of say the emission temperature and of the conductivity $k$
were so inaccurate that actually the ratio  $\frac{t_{spread}}{t_{radiation}}$ was
bigger than unity. In such a case a fit to $\xi$ would
be independent of the details of the calculation of this ratio and be given alone
as the correction factor to make the value
of $\frac{N\sigma}{M^2}$  in the critical pearl case (\ref{prediction}) be corrected to the experimental value (\ref{Freyf}).
This would lead to
\begin{eqnarray}
 \xi&=& \frac{\left .\frac{N\sigma}{M^2}\right |_{exp}}{\left .\frac{N\sigma}{M^2}\right |_{crit,\; all \rightarrow 3.5}}\nonumber\\
&=&\frac{(1.0\pm 0.2)10^{23} cm^2/kg^2}{2.0 \times 10^{21} cm^2/kg^2} =
50 \quad \hbox{(for $\phi_H =123$ GeV)}
\end{eqnarray}
and
\begin{eqnarray}
\xi
&=&\frac{(1.0\pm 0.2)10^{23} cm^2/kg^2}{1.0 \times 10^{21} cm^2/kg^2} =
100 \quad \hbox{(for $\phi_H=0$)}
\end{eqnarray}
Actually for $\xi=
50$ the time ratio is
\begin{eqnarray}
 \frac{t_{spread}}{t_{radiation}} &=& \frac{50^5}{2.7*10^8} = \frac{1}{8.6} \quad (\hbox{for} \;\phi_H=123 \;\hbox{GeV})
\end{eqnarray}
and for $\xi=
100$ it is
\begin{eqnarray}
\frac{t_{spread}}{t_{radiation}}&=& \frac{100^5}{4.2*10^9} = \frac{1}{0.42} \quad (\hbox{for} \;\phi_H=0).
\end{eqnarray}
So the factors by which we should have miscomputed e.g. the heat conductivity $k$ or
the fourth power of the emission temperature, in order
for this ``the ratio being bigger than unity'' situation to be realized, should be numbers
like 8.6 or 0.42 above.
We suggested above (\ref{1c5}) that the true emission temperature was about 1.5 times larger than
the line frequency 3.5 keV and that the conductivity should be reduced by a factor of 5 (\ref{emp}).
Each of these suggested improvements would increase the ratio $\frac{t_{spread}}{t_{radiation}}$ by a factor of 5,
so indeed this ratio is compatible with 1 for $\xi=50$ or $\xi=100$ in the respective cases.

If we included, say, one of these crudely estimated improvements by a factor of 5,
the $\xi$-values making the
ratio $\frac{t_{spread}}{t_{radiation}}$ just unity would be shifted from the $\xi_1$ values
above (\ref{xi1crit1}, \ref{xi1crit2}) to
\begin{eqnarray}
 \xi &=& \xi_1 = \sqrt[5]{\frac{2.7*10^8}{5}} = 35 \quad \hbox{(for $\phi_H=123\, \hbox{GeV}$)}\label{c15f123}\\
 \xi &=& \xi_1 = \sqrt[5]{\frac{4.2*10^9}{5}} =
61 \quad \hbox{(for $\phi_H=0$)}\label{c15f0}.
\end{eqnarray}
Written for the variable $\frac{\xi*10MeV}{\Delta V}$ stressed in the
next section,
these values giving the time ratio to be just unity are
$\frac{\xi*10MeV}{\Delta V}=\frac{10 MeV * 35}{135 MeV}= 2.6 $
and $\frac{\xi*10MeV}{\Delta V}=610/270 = 2.3$ respectively.
So, in the following section, we shall take  $\frac{\xi*10MeV}{\Delta V} > 2.4$
as the condition for $\frac{t_{spread}}{t_{radiation}} $
to be bigger than or equal to unity.

\subsection{Quantities to fit}

It turns out that  all of the four quantities whose values we would like
to use as tests of our model
depend on the same ratio $\frac{\xi* 10 MeV}{\Delta V}$
rather than on $\xi$ and $\Delta V$ separately, provided  $\frac{t_{spread}}{t_{radiation}} \ge 1 $
as assumed here. These quantities are:
\begin{itemize}
\item The frequency of the radiation 3.5 keV,
\item The intensity of the 3.5 keV radiation as given in (\ref{Freyf}),
\item The cube root of the surface tension $S^{\frac{1}{3}}$,
\item Theoretical value of $\xi$ and $\Delta V$ combined.
\end{itemize}

\begin{figure}
\begin{center}
\includegraphics[scale=0.4,angle=270]{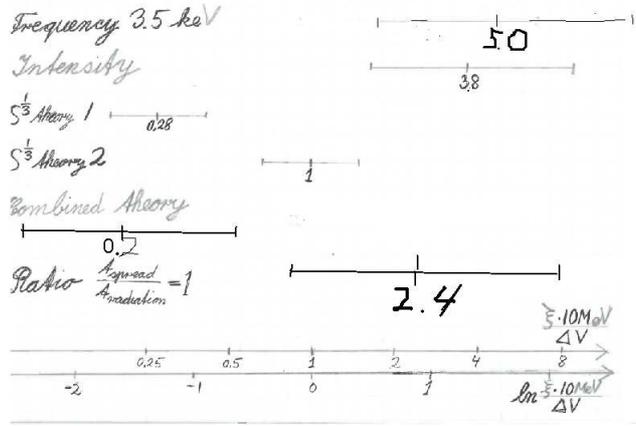}
\end{center}
\caption{The values of the ratio ${\frac{\xi* 10\ MeV}{\Delta V}}$
as needed for four constraints.
There are two experimental constraints from the frequency and intensity of the 3.5 keV radiation respectively
and two theoretical constraints in two versions corresponding to taking theory 1 or theory 2
for the tension.
We make the simplifying assumption that all energy from the surface contraction in a collision
gets emitted as 3.5 keV X-rays.
The sixth line ``Ratio $\frac{t_{spread}}{t_{radiation}}=1$'' represents the
condition for all the energy actually going to 3.5 keV radiation.
}
 \label{fig:figure1}
\end{figure}

We therefore plot the predictions for the ratio $\frac{\xi* 10MeV}{\Delta V}$ from these four quantities
in Figure \ref{fig:figure1}
with logarithmic uncertainties estimated crudely as seen in
Table \ref{table:table1}:

\begin{table}[h]
\begin{tabular}{|c|c|c|c|c|}
\hline
Name & $\frac{\xi* 10 MeV}{\Delta V}$ &$\ln{\frac{\xi* 10 MeV}{\Delta V}}$&Uncertainty&\\
\hline
Frequency ``3.5keV''& 5.0&1.61 &100\%&\\
Intensity $\frac{N\sigma}{M^2}$& 3.8 & 1.3 &90\%&\\
 $S^{1/3}$ theory 1)&0.28&-1.3&40\%&\\
 $S^{1/3}$ theory 2)&1&0&40\%&\\
Combined theory $\xi$, $\Delta V$&0.18&-1.7&100\%&\\
\hline
Ratio $\frac{t_{spread}}{t_{radiation}}$=1 & 2.4& 0.88& 80\%&l.b.\\
\hline
 \end{tabular}
\caption{Table of four theoretical predictions of the parameter
${\frac{\xi* 10 MeV}{\Delta V}}$ on which the quantities
happen to mainly depend.
The first column denotes the quantities for which we can provide a theoretical
or experimental value to be expected for our fit to that quantity.
The next column gives what these expected values need the
parameter combination ${\frac{\xi* 10 MeV}{\Delta V}}$ to be.
The third column is the natural logarithm of that required value for
the ratio ${\frac{\xi* 10 MeV}{\Delta V}}$, i.e.
$\ln{\frac{\xi* 10 MeV}{\Delta V}}$. The fourth column
contains crudely estimated uncertainties of the parameter
thus fitted counted in this natural logarithm.
In the last column we just marked the ratio $\frac{t_{spread}}{t_{radiation}}$
with l.b. to stress that it is only a lower bound and shall not be considered a
great agreement for our theory. }
\label{table:table1}
\end{table}

The entries in this table and figure were estimated as follows:
\begin{itemize}
 \item
 In our model the frequency of the radiation is equal to the energy gap $E_H$,
 which is inversely proportional to our parameter $\frac{\xi* 10MeV}{\Delta V}$.
Under the assumption
that $\Delta V = 135$ MeV and the pearl being of the smallest possible
size for stability, so that $\xi=1$, the homolumo gap (\ref{EHold1}) is
$E_H =
240$ keV.
So our parameter is required to take the value
\begin{eqnarray}
\frac{\xi* 10MeV}{\Delta V}& =&\frac{240 \ \hbox{keV}}{13.5*3.5\ \hbox{keV}} = 5.0
\end{eqnarray}
in order to predict a frequency of 3.5 keV.

Our calculations are almost just order of magnitude estimates and so crudely
we take the logarithm of the above determined value of our parameter to
have an uncertainty of order unity
so that
$\frac{\xi* 10MeV}{\Delta V}$=$  5.0^{+8.6}_{-3.2} =\exp(1.61 \pm 1)$.

\item
The intensity-related quantity $\frac{N\sigma}{M^2}$
is proportional to our parameter $\frac{\xi* 10MeV}{\Delta V}$.
Assuming that the time ratio $\frac{t_{spread}}{t_{radiation}} \ge 1 $,
the value of $\xi$ is given
by the ratio of the experimental value (\ref{Freyf}) to our predicted value
(\ref{prediction}) for a critical size pearl with
 $\Delta V = 135$ MeV and all
the energy released from the contraction of the collision surface going into the 3.5 keV line.
This leads to the following value of our parameter:
\begin{equation}
\frac{\xi* 10MeV}{\Delta V} = \frac{(1.0 \pm 0.2)*10^{23}\ \hbox{cm$^2$/kg$^2$}}{13.5 *2.0 * 10^{21}\  \hbox{cm$^2$/kg$^2$}} = 3.8
\end{equation}

The uncertainty here is likely to be dominated by the lack of good
knowledge as to the clumpiness of the dark matter, see the column
``boost''   in the table in  section \ref{table0}. Including also
the uncertainty in our crude estimate of the energy released
from the contraction of the pearl surfaces during the collision
etc., we estimate a total uncertainty of about 90\%  for this restriction on
$\ln{\frac{\xi* 10 MeV}{\Delta V}}$.

\item $\sqrt[3]{S}$ theory:
\begin{itemize}
\item{Tension theory 1.)}

In the appendix
\ref{theory1} we calculate
an upper limit for the tension (\ref{Slimit}) $S \le (140\ \hbox{GeV})^3$
mainly given in terms of the parameters of the Higgs
field. Since this is an upper limit, we have taken
$S = (100\ \hbox{GeV})^3$ as a realistic estimate of the tension.
Now $S^{1/3}$ is inversely proportional to the
parameter ${\frac{\xi* 10 MeV}{\Delta V}}$
and its value for a critical sized pearl
is given by \cite{Tunguska}:
\begin{equation}
S_{crit}^{1/3} = \left(\frac{M}{24\pi^5 m_N}\right)^{1/9} \Delta V.
\end{equation}
For a pearl of mass $ M = 1.4*10^8$ kg and with $\Delta V = 135$ MeV
this tension for a critical sized pearl is determined to be
$S^{1/3}= 380$ GeV.
This leads to the requirement
\begin{eqnarray}
\frac{\xi* 10 MeV}{\Delta V} =\frac{380\ \hbox{GeV}}{13.5* 100\ \hbox{GeV}}
= 0.28.
\end{eqnarray}
One here a priori calculates $S$ but the parameter $\frac{\xi* 10 MeV}{\Delta V}$
is inversely linearly related to the cubic root of this quantity,
namely to $S^{\frac{1}{3}}$. So the a priori typical uncertainty
gets reduced by a factor 3, and we end up estimating the uncertainty
on the extracted value
of $\ln \frac{\xi 10* MeV}{\Delta V}$ to be rather about one third of the ``usual''
100\%, taken to be 40\%.

\item{Tension theory 2.)}

In the appendix \ref{RA}
we propose the existence of a third low energy vacuum which, according
to the multiple point principle, is degenerate with the other two vacua. In this way we
are naturally led in appendix \ref{theory2} to a smaller theoretical value $S \sim (30\ \hbox{GeV})^3$.
This value $S^{1/3}\sim 30 $ GeV leads to the requirement
\begin{eqnarray}
\frac{\xi* 10 MeV}{\Delta V} =\frac{380\ \hbox{GeV}}{13.5*30 \ \hbox{GeV}} \sim 1.
\end{eqnarray}
with a similar uncertainty to that of Tension theory 1.
\end{itemize}
\item
In the fifth  row of Table \ref{table:table1} we give the value
of this parameter ${\frac{\xi* 10 MeV}{\Delta V}}$ simply obtained from
our best theoretical ideas for $\xi$ and $\Delta V$ separately:
\begin{itemize}
\item $\xi = \sqrt{4\pi}*2^{\frac{4}{9}} =4.8$, see (\ref{xitheory}) below.
\item In the appendix \ref{smooth} we argue for $\Delta V=270$ MeV being the
most likely value.
\end{itemize}
Combining these separate values leads to ${\frac{\xi* 10 MeV}{\Delta V}}=
\frac{4.8*10MeV}{270MeV}=0.18 $.
Taking the logarithmic uncertainty on each of these values to be 70\%,
we obtain an overall uncertainty of 100\% on $\ln \frac{\xi 10* MeV}{\Delta V}$.

\item In the last row in the table - separated by a line - we
present the condition that the ratio $\frac{t_{spread}}{t_{radiation}}$
be bigger than or equal to unity, in order that the fitting used for
the other quantities becomes relevant.
The value  $\frac{\xi *10MeV}{\Delta V} =
2.4 $ is in fact the lower
limit needed to ensure that $\frac{t_{spread}}{t_{radiation}} \ge 1$.
However it turns out that some of the fits considered below tend to violate this
lower limit. So, in practice, we replace the inequality by an equality
with an estimated uncertainty of 80\% on $\ln \frac{\xi 10* MeV}{\Delta V}$.

However, this last
lower limit is not a genuine prediction, but rather
a condition it turned out we needed to impose and
presumably represents a warning that something is unexpected in our model.

\end{itemize}

\subsection{The Various Fits} \label{various}
Using the data put forward in the table we here present four fits using
selected amounts of these data:

\begin{itemize}
\item{\bf  Tension theory 1) Fit}

The a priori simplest fit to make would be to use the theory 1) for
the value of $S^{\frac{1}{3}} = 100$ GeV, which just uses the simplest
assumption about the field expectation values for the vacua of the two phases
``present'' and ``condensate'' vacua.
The average of the parameter fitted becomes
\begin{eqnarray}
<\ln\frac{\xi*10 \: MeV}{\Delta V}>&=&
-0.47\pm 0.3\\
\Rightarrow \quad <\frac{\xi*10 \: MeV}{\Delta V}> &=&
\exp(
-.47) = 0.62^{+0.2}_{-0.2}\\
\chi^2&=&
16.8\hbox{ (for 4 degrees of freedom)}\nonumber
\end{eqnarray}
But this fit is
not so good in as far as the tension and the measured X-ray line frequency and intensity
each deviate from their fitted value by 2 standard deviations.





\item{\bf Tension theory 2) Fit}

If we accept the existence of another minimum of the effective potential
as a function of the fields for the bound state $\phi_F$ and the Higgs field $\phi_H$
degenerate with the two minima associated to the ``present''
and ``condensate'' vacua, then we are led to a lower estimate for the tension
$S^{1/3} \sim 30$ GeV.
A smaller $S$ means it can be fitted by a larger value of $\frac{\xi*10MeV}{\Delta V}$
than for theory 1).
In this theory 2) we get :
\begin{eqnarray}
<\ln\frac{\xi*10 \: MeV}{\Delta V}>&=&
0.26\pm 0.3\\
\Rightarrow \quad <\frac{\xi*10 \: MeV}{\Delta V}>&=& \exp(
0.26) = 1.3\pm 0.4\\
\chi^2 &=&
8.02\hbox{ (for 4 degrees of freedom)}
\end{eqnarray}
This fit is  more satisfactory
with respect to the tension which now only deviates
by 0.7 st.d.,
but the ``combined theory" now deviates by almost 2
standard deviations.

This tension theory 2) fit including theoretical and empirical constraints gives the following overall
fitted values for the experimental
quantities
\begin{eqnarray}
\hbox{The frequency predicted = $E_H$ }
&=&
14^{+24}_{-9} \, \hbox{keV},
\end{eqnarray}
and
\begin{eqnarray}
\hbox{The intensity predicted =} \, \frac{N\sigma}{M^2}
&=&
3.5^{+5.2}_{-2.1}*10^{22}\, \hbox{cm$^2$/kg$^2$},\label{in1}
\end{eqnarray}
which are to be compared with the line frequency of 3.5 keV and the observed intensity
\begin{eqnarray}
 \left.\frac{N\sigma}{M^2}\right |_{exp} &=& (10\pm 2 )*10^{22}\, \hbox{cm$^2$/kg$^2$}
\end{eqnarray}

However there is clearly a tension between the theoretical constraints, which favour
lower values of the parameter $\frac{\xi*10 \: MeV}{\Delta V}$, and the empirical
constraints which favour larger values. We will now consider fits to the theoretical
and empirical constraints separately.

\item{\bf Theoretical Fit}

Using only the theoretically predicted quantities - the tension theory 2), the combined theory $\xi$ and $\Delta V$, and the restriction $\frac{t_{spread}}{t_{radiation}} =1$ -
so that we get what
can be considered as a purely theoretical
fitting, we obtain
\begin{eqnarray}
< \left. \ln \frac{\xi*10 \: MeV}{\Delta V} > \right |_{theory}&=&
-0.04 \pm 0.34\\
\Rightarrow \quad \left . <\frac{\xi*10 \: MeV}{\Delta V}>\right |_{theory}&=&\exp(
-0.04) = 0.96^{+0.4}_{-0.3}
\end{eqnarray}
\begin{equation}
\chi^2 =
4.09 \hbox{ (for 2 degrees of freedom)}
\end{equation}

Using our estimated logarithmic uncertainties and the
purely theoretical value of $\left. \ln \left (\frac{\xi*10 \: MeV}{\Delta V}\right )\right |_{theory}=
-0.04$
we obtain the following values for the experimental quantities:
\begin{eqnarray}
 \hbox{The frequency predicted $=E_H$}
 &=&
18^{+31}_{-11} \, \hbox{keV} \label{fr}
\end{eqnarray}
and
\begin{eqnarray}
\hbox{The intensity predicted =} \, \frac{N\sigma}{M^2}
&=&
2.5^{+3.6}_{-1.5}*10^{22}\, \hbox{cm$^2$/kg$^2$}. \label{in1a}
\end{eqnarray}
These values are to be compared with the line frequency of 3.5 keV and the observed intensity
\begin{eqnarray}
 \left.\frac{N\sigma}{M^2}\right |_{exp} &=& (10\pm 2 )*10^{22}\, \hbox{cm$^2$/kg$^2$}
\end{eqnarray}

The errors in the above predictions are dominated by the uncertainties estimated for
the calculations of the homolumo gap and the rate of X-ray production respectively, so that
the uncertainty of only 35 \% in the parameter $\frac{\xi*10MeV}{\Delta V}$ is almost negligible.
But we did indeed include even this little uncertainty in the results  presented in (\ref{fr}, \ref{in1a}).

These numbers (\ref{fr}) and (\ref{in1a}) can be considered theoretical predictions
only based on the rate of Tunguska-like events in our model. They agree with the data in order of magnitude.


{\bf Empirical Fit}

Using only the experimentally observed quantities - the X-ray line frequency
and intensity - so that we obtain purely a fit to the empirical data, we obtain
\begin{eqnarray}
<\ln \frac{\xi*10MeV}{\Delta V}> &=& 1.44 \pm 0.67 \\
\Rightarrow <\frac{\xi*10MeV}{\Delta V}>&=& 4.2^{+4.0}_{-2.0}
\end{eqnarray}
\begin{equation}
\chi^2 = 0.05 \hbox{ (for 1 degree of freedom)}
\end{equation}

Corresponding to this fit
we have
\begin{eqnarray}
\hbox{The frequency predicted} = E_H
&=& 4.2^{+7.2}_{-2.7} \, \hbox{keV},\\
\hbox{The intensity predicted =} \, \frac{N\sigma}{M^2}
&=& 11.0^{+19}_{-7}*10^{22}
\frac{cm^2}{kg^2}.
\end{eqnarray}
These values are in remarkably good agreement with 3.5 keV and
with $(10 \pm 2)*10^{22} \frac{cm^2}{kg^2}$
respectively. This empirical fit also satisfies the lower bound
$\frac{\xi*10MeV}{\Delta V} > 2.4$ coming from the condition that the
time ratio $\frac{t_{spread}}{t_{radiation}}$ be greater than one. However the
tension corresponding to the fit
\begin{equation}
S^{1/3} = 6.7^{+3.3}_{-2.2} \, \hbox{GeV}
\end{equation}
is significantly smaller than the theoretical values of 100 GeV or 30 GeV
from tension theory 1) and 2) respectively.


\end{itemize}

The empirical fit above shows that our pearl model of dark matter
is very successful in reproducing the two experimental quantities
concerning the 3.5 keV X-ray radiation: the frequency and the intensity
fitted to the various clusters and galactic center etc.
However our theoretical estimates of the model parameters  are only consistent
with the empirical fit within an order of magnitude.
In particular the low  value of the surface tension $S^{1/3} \sim 7 $ GeV obtained
is surprisingly small from a theoretical point of view.

\section{Some Further Corrections
}\label{further}

There are two smaller improvements which we might include, but which were left
out of the discussion above so as not to overcomplicate it:

\begin{itemize}
\item Provided that $\frac{t_{spread}}{t_{radiation}} <1$ then after the time $t_{spread}$ the hot spot has spread to the whole
pearl and the 3.5 keV radiation stops and gets replaced by the
thermal radiation from the now getting hotter and hotter surface of
the pearl. However,  after some further time the whole pearl gets
so cold that the temperature at the surface of the pearl again reaches
down into the 3.5 keV range, and then the emission of 3.5 keV radiation
is reinstated.  Now the hot spot, which is not so hot as the
original one but still much hotter than 3.5 keV in the interior,
begins to contract while emitting 3.5 keV radiation from the
boundary along which the temperature is of that order. This boundary
contracts as time then goes on more and more; first at the end, when
even the temperature in the center of the hot spot has cooled,
the emission of 3.5 keV radiation stops forever.

At most  we should expect that this effect of the contraction era for the hot spot
should give an amount of 3.5 keV radiation about the same as the expansion era. So
a factor of 2 increase in the amount of 3.5 keV radiation would be an upper limit
for this effect of a re-contraction. In spite of the average temperature being
smaller at the contraction of the 3.5 keV emitting surface, the temperature at the
emitting surface remains the same and thus the emission per unit time only depends on the
area of the emitting surface, which we just approximate by the surface area of the pearl.

We conclude that the fraction of the emitted energy appearing as 3.5 keV
radiation should, because of this correction, be increased by a factor between 1 and 2.


\item The border of the hot spot at which the 3.5 keV radiation is emitted
and which thus defines the size of this hot spot which really counts
for our calculation is not, as we used it, the radius $|\vec{r}|$ at which the
expression in the exponent of equation (\ref{kerneln}) is just unity. It should really be
the distance at which the temperature has got a value of the order
of the homolumo gap $\sim$ 3.5 keV. This means that the exponent
should rather than just be of the order $e^{-1}$ be of the order of the
ratio of this ``low'' 3.5 keV temperature divided by the typical
temperature from the start of the spread of the hot spot, or by some
average of that temperature. Now the initial temperature is of the
order of say (\ref{temp1}, \ref{temp2})
6 MeV (for $\phi_H = 123$) GeV, or 12 MeV (for $\phi_H =0$) or we use some average a bit smaller, and the low temperature
at which the 3.5 keV radiation gets emitted is of the order
of 3.5 keV. So the value of the quantity in the exponent should
rather be taken to be
$\ln(\frac{6\, MeV}{3.5\, keV})  = 7.4$ (for $\phi_H=123$ GeV)
or
$\ln(\frac{12\, MeV}{3.5\, keV})  = 8.1$ (for $\phi_H=0$).
This means that the true value for the spreading time $t_{spread}$,
and hence for the ratio $\frac{t_{spread}}{t_{radiation}}$, should be
reduced by a factor of
$7.4$ or $8.1$, from the above estimated values.

\end{itemize}

Taking say for the first of the above corrections an increase by a factor 1.3 in the amount of
3.5 keV radiation per collision, we get a total  decrease from our two corrections here by a factor of
$\frac{7.4}{1.3}=5.7$ for $\phi_H = 123$ GeV and $\frac{8.1}{1.3} =6.2$ for $\phi_H =0$.
So to compensate for such an extra factor  we must
increase our fitted $\xi$ values
by a factor of
\begin{equation}
correction \ factor =
\sqrt[6]{5.7} = 1.3 \quad (\hbox{for} \ \phi_H=123 \ \hbox{GeV}) \label{xxi123}
\end{equation}
and
\begin{equation}
correction \ factor =
\sqrt[6]{6.2} = 1.4 \quad (\hbox{for}\ \phi_H=0).\label{xxi0}
\end{equation}

\section{Speculative estimate of $\frac{R}{R_{crit}}$.}
\label{thxi}
In our previous article \cite{Tunguska} we took it as a likely hypothesis that
the size of the pearls would be close to the critical
size at which the pressure would be
so big that the nucleons inside the pearl
would just be about to be spit
out. It is of course likely that, on the average, the actual
size  of the pearls produced in the
early Universe would be close to this critical value.
Pearls of smaller size will of course disappear because they
will collapse. If thus the production of the pearls is somehow
biased towards rather small pearls the average size will
come close to the critical one.

Now, however, we must imagine that the pearls in the
creation era are not yet perfectly spherical but rather
have highly deformed shapes. In such a situation
the curvature of the skin around the pearl is not the same
all around, but varies from place to place.
Now whenever the curvature reaches the value of the critical
size spherical pearl the nuclei may start to be spit out.
Once the spit out starts the pearl gets volume-wise smaller
and thus once such a collapse has started it may very easily
come to continue. Thus for a pearl of the formally
just barely stable size, the smallest fluctuation
in the surface relative to sphericity would cause its collapse.
Thus the minimally required size of a pearl would be such
that the fluctuations in the curvature
as you go around the pearl surface never cause the curvature to reach the true
critical value, so that the curvature remains
less than the critical one all around the pearl. A situation in which the
curvature varies statistically around the pearl and
does not meet the critical value in more than one point should
effectively replace the critical radius pearl in our earlier considerations.
That is to say that the
statistical distribution of pearl sizes should begin rather
from this size in which the critical curvature is only met in one
point around the surface. Then from there on the distribution
should fall off in some way hoped to be biased towards small
pearls.

Can we obtain at least a very crude estimate of how big compared
to a true critical size spherical pearl the statistical
non-spherical randomly shaped pearl should be? We could
say that if the random variation of the curvature
varies on a scale given by the radius of the pearl over the
pearl surface, then it would be effectively as if there were
$4\pi$ approximately independent regions of the
surface. Each of these regions might take its own
curvature - say by dimensional argument of the
same order as the pearl average. This would mean
that we would expect the spread in the curvature, as you
go around the pearl, to be of a similar order as the
average  say of $4\pi$ independent variables with
each having a spread of one unit in terms of the
pearl average curvature. To look for the
effective collapse size, which we now estimate, we
may take the unit curvature here to be the absolute
collapse curvature for the exactly spherical pearl.

Now we simply say: The average of $4\pi$ independent
variables with spread 1 has spread $\sqrt{4\pi}$.
We must ensure that this fluctuating curvature
shall not reach the critical size, and that we do by making the
pearl $\sqrt{4\pi}$ times larger than the genuine
critical size. Thereby namely the curvatures in the
fluctuation around the surface get scaled down by a
factor $\sqrt{4\pi}$ and just becomes unity.

So we argue that the scaling ratio $\xi$ of the average
pearl size relative to the genuine critical size
should be at least $\sqrt{4\pi}$ to just avoid
collapse.

However, as we estimated in appendix B2 of \cite{Tunguska},
there will in addition be some excess in size over the effectively
critical size because of fluctuations in the formation of the pearls .
The latter increase in the expected size depends on how biased
the size is towards small sizes. But it means that we expect at least
\begin{equation}
 \xi \ge \xi_{crit} = \sqrt{4\pi}\label{xicrit}.
\end{equation}
In fact we predicted that the statistical distribution of pearl sizes would give
\begin{equation}
 \frac{R_{median}}{R_{crit}} = 2^{4/9}\approx 1.4
\end{equation}
for the ratio of the median radius $R_{median}$ to the critical radius $R_{crit}$.
This value should be multiplied by the critical size parameter $\xi_{crit}$ above (\ref{xicrit}).
Thus the final prediction for the ratio $\frac{R}{R_{crit}}$ becomes
\begin{equation}
 \xi = 2^{4/9}*\sqrt{4\pi} \approx 5. \label{xitheory}
\end{equation}

\section{Conclusion}

We have developed our earlier published model \cite{Dark1,Dark2,Tunguska,Supernova} for dark matter being balls or pearls
of centimeter size and a mass of the order of 140000 ton so as to investigate whether
our pearls can deliver
the controversial 3.5 keV X-ray line. Our picture is that this 3.5 keV radiation
appears from collisions of pairs of our dark matter pearls; the pearls unite in the collision
and by the contraction of their skin liberate so much energy $E_S$
as heat in the ball that it can produce sufficient 3.5 keV X-rays
to reproduce the observed radiation in this line.


In our model we identify the photon energy 3.5 keV
with the homolumo energy  gap in pearl material.
We take it, that there is very generally a homolumo gap effect in for instance glassy
materials - as we can suspect our pearl material to be - consisting in that the nuclei and the
electrons will adjust to arrange the  empty (single) electron states to increase in energy,
while the filled ones will sink in energy. By such an arrangement the single electron energy of the Fermi sea
is lowered.
Thereby a gap appears between the empty and the filled levels and that is called
the homolumo gap.

The existence of a homolumo gap
means that strictly speaking the material is
an insulator or a semi-conductor in spite of it being in first approximation, when the
homolumo gap is ignored, a metal. With such an energy gap, it can easily happen that
excited electrons get collected just above this gap and holes of missing electrons just below.
When such holes annihilate with these excited electrons, radiation with energy
essentially equal to the homolumo gap will be emitted. We can therefore expect
radiation from say our dark matter pearls with a frequency just equal to the
homolumo gap in frequency.

From the  analysis by Cline and Frey \cite{FreyCline} of the observed intensity of
the 3.5 keV X-ray line, we extracted the result
\begin{equation}
\left. \frac{N\sigma}{M^2}\right |_{exp} = (1.0\pm 0.2)*10^{23}\, \hbox{cm$^2$/kg$^2$}. \label{exp-intensity}
\end{equation}
Here N denotes the number of 3.5 keV photons emitted in a pearl collision,
$\sigma = \pi (2R)^2$ is the collision cross section and M is the pearl mass.
If all the energy $E_S$ released due to the contraction of the skin around the pearls
after a collision is radiated as 3.5 keV photons
\begin{equation}
N = \frac{E_S}{3.5 \, \hbox{keV}}.
\end{equation}
However, if the time $t_{spread}$ for the hot spot produced in the collision to spread
over the pearl is less than the time $t_{radiation}$ it takes to radiate out all the
energy $E_S$ into 3.5 keV X-rays, we estimate that N is reduced by a factor
$\frac{t_{spread}}{t_{radiation}}$.

In the old theory used in our Tunguska-paper \cite{Tunguska} we used the
approximation or theoretical assumption that the pearls had just such a size that
they were on the borderline of collapsing, because the the material
was just about to be pressed out of the pearls across the $\Delta V$ potential per
nucleon.
However this hypothesis of exact borderline stability is unrealistic. So we
introduced a parameter $\xi = \frac{R}{R_{crit}}$
denoting the radius of the actual pearl relative to the radius of a critical
size pearl which is
just barely stable. The radius of a critical size pearl with the
mass $M = 1.4 * 10^8$ kg estimated from the Tunguska event \cite{Tunguska} and
with $\Delta V = 135$ MeV is $R_{crit} = 0.05$ cm.

Now we first summarize the results for a critical sized pearl and then
for a pearl of arbitrary radius.

\begin{itemize}
\item{\bf Pearls of critical size}



 Using the parameters in Table \ref{T} (with $\Delta V = 135$ MeV) as determined from
the Tunguska event and estimates performed in earlier works, before we
studied the 3.5 keV line, we calculated by use of the Thomas Fermi approximation or
just by dimensional arguments the homolumo gap value to be
\begin{eqnarray}
E_H &=& 240 \, \hbox{keV}.
\end{eqnarray}
This is a factor of 70 too large.

For critical sized pearls we obtain a very small time ratio
\begin{eqnarray}
 \frac{t_{spread}}{t_{radiation}}
&=&\frac{1}{2.7*10^{8}}*\exp(\pm 400 \%) \quad  (\hbox{for }\phi_H=123 \; \hbox{GeV})
\end{eqnarray}
and we end up with the prediction for the intensity related expression
(\ref{prediction}) becoming
\begin{eqnarray}
 \left. \frac{N\sigma}{M^2}\right |_{th. \: crit} &=&
7.4 *10^{12}\exp(\pm 400\%)\, \hbox{cm$^2$/kg$^2$}.
\end{eqnarray}
This is ten orders of magnitude less than the experimental
value (\ref{exp-intensity}).

We conclude that critical sized pearls with $\Delta V =$ 135 MeV or 270 MeV
are inconsistent with the observed frequency and intensity of the 3.5 keV line.

However the ratio $\frac{t_{spread}}{t_{radiation}}$
giving the fraction of the energy appearing as the 3.5 keV line radiation
is very sensitive to the new parameter $\xi$ or $\frac{\xi* 10 MeV}{\Delta V}$,
which we now introduce. We can only hope for a good fit, if this parameter is chosen to
make the time ratio $\frac{t_{spread}}{t_{radiation}}$ very close to or greater
than unity.

\item{\bf Pearls of arbitrary radius}

We here consider a pearl of radius $R = \xi * R_{crit}$. It turned out
that all the important features of our pearls
depend only on the ratio of this new
parameter $\xi$ relative to
the already used parameter $\Delta V$ (denoting the potential barrier for a nucleon to leave
the pearl). Since in \cite{Tunguska} we used $\Delta V =$ 10 MeV, this
composed parameter $\frac{\xi* 10 MeV}{\Delta V}$ was equal to unity in
the old calculations. It follows that
\begin{equation}
R = \frac{\xi* 10 MeV}{\Delta V}*0.67\, \hbox{cm} \quad \hbox{and} \quad
\quad S^{1/3} = \frac{28\, \hbox{GeV}}{\frac{\xi* 10 MeV}{\Delta V}}.
\end{equation}
The condition that the time ratio $\frac{t_{spread}}{t_{radiation}}$ be greater than unity and hence
not suppressing the intensity of the 3.5 keV line is
\begin{equation}
\frac{\xi* 10 MeV}{\Delta V} > 2.4. \label{lb2}
\end{equation}

In the present work we calculated a value for this parameter by fitting the
frequency and observed intensity (\ref{exp-intensity}) of
the 3.5 keV line. This empirical fit gave the value $\frac{\xi* 10 MeV}{\Delta V} = 4.2^{+4.0}_{-2.0}$, which
satisfies the bound (\ref{lb2}). Corresponding to this fitted value of our parameter we have
\begin{equation}
E_H = 4.2^{+7.2}_{-2.7} \, \hbox{keV} \quad \hbox{and}
\quad \frac{N\sigma}{M^2} = 11.0^{+19}_{-7}*10^{22} \,
\frac{cm^2}{kg^2}.
\end{equation}
Also we obtain the following values for the radius and tension of the pearls
\begin{equation}
R = 2.8^{+2.7}_{-1.3} \, \hbox{cm} \quad \hbox{and}
\quad S^{1/3} = 6.7^{+3.3}_{-2.2} \, \hbox{GeV}.
\end{equation}
This excellent fit to both the experimental quantities  shows that our pearl model
is successful in explaining the origin of the 3.5 keV line.

We have also made theoretical estimates for the surface tension and the
parameters $\xi$ and $\Delta V$, which turn out to be in some tension with
the above empirical fit. The most natural estimate for the surface tension
is $S^{1/3} \sim 100$ GeV, while the more complicated tension theory 2) of appendices
\ref{RA} and \ref{theory2} including the existence of a speculated extra
vacuum gives $S^{1/3} \sim 30$ GeV. These values are
significantly larger than $S^{1/3} = 6.7$ GeV obtained from the fit.
Similarly the theoretical estimate (\ref{xitheory}) of $\xi \sim 5$
gives a pearl radius R = 0.25 cm for $\Delta V = 135$ MeV or
R = 0.125 cm for $\Delta V = 270$ MeV; an order of
magnitude smaller than the fitted value R = 2.8 cm.

In section \ref{various} we have considered various fits in which we have included
our theoretical estimates as quantities to be fitted by our parameter
$\frac{\xi* 10 MeV}{\Delta V}$. These fits tend to violate the condition
(\ref{lb2}) and so we included the lower bound as a quantity in the fits.
In this way it is possible with tension theory 2) to obtain an acceptable order
of magnitude fit with $\frac{\xi* 10 MeV}{\Delta V} = 1.3 \pm 0.4$. However we
have most confidence in our empirical fit above and feel that we need a better
understanding of the surface tension and radius of our pearls.

\end{itemize}

\section{R\'{e}sum\'{e} of Present Fit Parameters}

In the table below we now present the parameters of our model picture of the Tunguska particle
as a pearl of a new type of vacuum with a bound state condensate, filled
with ordinary white dwarf-like matter.
In column 4 we give the results for a critical pearl on the borderline of stability
as used in our previous work \cite{Tunguska}
except that we correct the $\Delta V =10$ MeV  used there into
$\Delta V = 135$ MeV as given by the best estimate of the  Higgs nucleon interaction \cite{gNN}.
In column 5 we present a fit with one more parameter i.e. the ratio of the parameter
$\xi$, denoting the pearl's radius divided by the radius of a critical pearl, and
the potential difference $\Delta V$ for a nucleon to pass through the skin of the pearl (multiplied by 10 MeV)
$\frac{\xi*10MeV}{\Delta V}$. The values in column 5 correspond to the value $\frac{\xi*10 MeV}{\Delta V}=4.2$
for our "empirical fit" to the X-ray line frequency and intensity.
\begin{center}
\begin{longtable}{ | l | l | l | l |l|}
\hline
Nr. &Name& symbol&old $\frac{\xi* 10Mev}{\Delta V}=\frac{1}{13.5}$ &new  $\frac{\xi* 10MeV}{\Delta V}=4.2$\\
\hline
\endfirsthead
\hline
Nr. &Name& symbol&old $\frac{\xi* 10Mev}{\Delta V}=\frac{1}{13.5}$ &new  $\frac{\xi* 10MeV}{\Delta V}=4.2$\\
\hline
\endhead
\endfoot
\endlastfoot
  1.  &Time Interval of impacts & $r_B^{-1}$ & 200 years&kept\\
\hline
 2.&   Rate of impacts& $r_B$& $1.5 *10^{-10}\ \hbox{s}^{-1}$&    \\ \hline
 3.&   Dark matter density
& $\rho_{halo}$ &0.3 GeV/cm$^3$ &kept\\
&in halo&&&\\
\hline
 4.&   Dark matter
solar system & $\approx 2\rho_{halo}$ &0.6 GeV/cm$^3$&kept   \\ \hline
5.& 
Typical speed of ball& $v$& 160 km/s&kept \\ \hline

 6.&   Mass of the ball & $m_B$ & $1.4*10^8$ kg & kept  \\
&&&=140000ton&\\
&&&$=7.9*10^{40}\frac{{keV}}{c^2}$&\\
    \hline

7.& Kinetic energy of ball & $T_v$ & $1.8*10^{18}$ J=&  
 \\
&&& 430 Mton TNT&\\
\hline
8. &Energy observed,  Tunguska& $E_{Tunguska}$ & $(4-13)*10^{16}$ J=&
\\
&&& 10 - 30 Mton TNT&\\ \hline
9.& Potential shift
& $\Delta V$ & 135 MeV  &(270 MeV) \\
& between vacua &&&\\
\hline
10.&$\sqrt[3]{\hbox{tension}}$(fit) & $S^{1/3}$&380 GeV
& 6.7 GeV\\
 \hline
11.1&$\sqrt[3]{\hbox{tension}} $(condensate) 1)& $S^{1/3}$ & 16 GeV &100 GeV \\
11.2&$\sqrt[3]{\hbox{tension}} $(condensate) 2)&&&30 GeV\\
\hline
12. &Ball density& $\rho_B$ & $2.5*10^{17}\frac{ kg}{m^3}
$ &$1.4*10^{12}\frac{kg}{m^3}$ \\
\hline
13.&Radius of ball & $R$ & 0.05 cm&2.8 cm \\
\hline
14.& homolumo gap&$E_H$&
240 keV&4.2 keV\\
\hline
15. &Frequency(obs.) &$3.5\ \hbox{keV}$&3.5 keV& - \\
\hline
16 &Released energy&$E_S$&
$5.4 \% Mc^2=$&0.095 \% $Mc^2=$\\
&&&=$
4.3 *10^{39}\ \hbox{keV}$&$7.6*10^{37}\ \hbox{keV}$\\
\hline
17.&\# 3.5's if all$\rightarrow$ 3.5&$N_{\hbox{all$\rightarrow$ 3.5}}$
&
$1.24 *10^{39}$&$2.2 *10^{37}$\\
&&$=\frac{E_S}{3.5\ keV}$&&\\
\hline
18. & Spreading time & $t_{spread}$&
$5.2*10^{-3}$s&
17 s \\
\hline
19. & Radiation time&$t_{radiation}$&
$1.4*10^6$s&7.7 s\\
19 b.&with 1.5: &&
$2.8*10^5$s& 1.5 s\\
\hline
20. & Ratio&$\frac{t_{spread}}{t_{radiation}}$&$\frac{1}{2.7*10^8}$&$2.2 \rightarrow 1$\\
20 b.&with 1.5:&&$\frac{1}{
5.4*10^7}$&$11 \rightarrow 1$\\
\hline
21.& \# 3.5's&$N=\frac{t_{spread}}{t_{radiation}} *$&
$4.6*10^{30}$&$
2.2*10^{37}$\\
&&$*N_{\hbox{all$\rightarrow$3.5}}$&&\\
21b.&with 1.5: &&
$2.3*10^{31}$&$2.2*10^{37}$\\
\hline
22. &cross section, balls&$\sigma=\pi(2R)^2$&
0.031$ cm^2$& 100 $cm^2$\\
\hline
23. & cross section per $\gamma$&$N_{all\rightarrow 3.5}\sigma$&
$3.8*10^{37}cm^2$
&$2.2 *10^{39}cm^2$\\
& as if all$\rightarrow$ 3.5&&&\\
\hline
24.&cross section per $\gamma$&$N\sigma$&$
1.4 *10^{29}cm^2$&$
2.2*10^{39}cm^2$\\
&(with time ratio)&&&\\
24b.&with 1.5:&&
$7.1*10^{29}cm^2$&$
2.2*10^{39}cm^2$\\
\hline
25.& $\sigma$ per $\gamma$ per $M^2$&$\frac{N\sigma}{M^2}|_{all\rightarrow 3.5}$&
$2.0*10^{21}\frac{cm^2}{kg^2}$&$11*10^{22}\frac{cm^2}{kg^2}$\\
& (as if all $\rightarrow$ 3.5)&&&\\
\hline
26. & $\sigma$ per $\gamma$ per $M^2$&$\frac{N\sigma}{M^2}=$
&$
7.2*10^{12}\frac{cm^2}{kg^2}$&$
11*10^{22}\frac{cm^2}{kg^2}$\\
&(with time ratio)&$\frac{t_{spread}N_{all\; \rightarrow \; 3.5}\sigma}
{t_{radiation}M^2}$&&\\
26b.& with 1.5:&& $
3.6*10^{13}\frac{cm^2}{kg^2}$&$
11*10^{22}\frac{cm^2}{kg^2}$\\
\hline
27. &Fit to Cline and Frey&$\frac{N\sigma}{M^2}$& $(1.0\pm 0.2)10^{23}
\frac{cm^2}{kg^2}$&- \\
& (including Boost corr.)&&&\\
\hline
28.& Radius/critical R&$\xi = R/R_{crit}$& 1&
57 \\
& (fitted)&&&\\
28b.&with$<\phi_H>=0$&$\xi$& -&
113 \\
\hline
29.& Radius/ critical R& $\xi$=  &- &$2^{4/9}\sqrt{4\pi}$\\
& (speculated)&$2^{4/9}\sqrt{4\pi}$ &&=4.82\\
\hline
30.&Heat  conductivity&$k=\frac{c^2p_f^2}{55\alpha}$&
$2*10^5\frac{MeV^2}{c}$&
60 $\frac{MeV^2}{c}$\\
\hline
\end{longtable}
\end{center}

\subsection{Review of Definitions and Explanation of the Table}
Let us here shortly review the concepts given and explain the
table:
 The second column contains the short name of the
quantity given in our model, and the third column is the
formula expression for it. The fourth and the fifth columns
contain suggested numerical order of magnitude values for the
quantity in question: The fourth column gives the value obtained
with the old numbers from our previous publication
\cite{Tunguska} updated with $\Delta V =135$ MeV, so that what could be gotten from
these numbers could be considered in some sense ``pre''diction.
These numbers were based in some cases on the hypothesis that
the {\bf size of the typical pearl is such that it is just
on the borderline of stability towards collapsing by the
matter/nuclei inside being spit out under the pressure.}
However further investigation suggests this hypothesis is not
realistic and the actual radius $R_{actual}$ of a pearl
is instead  taken to be a fitting parameter $\xi$ times
larger than the borderline radius $R_{crit}$.
Then we actually use the parameter
$\frac{\xi*10 MeV}{\Delta V}$ to fit both
the observed frequency and intensity of the X-ray line.
The fitted value $\frac{\xi*10 MeV}{\Delta V} =
4.2$ is used in the fifth column
of the table.



Now a short review of the rows in the
table:

\begin{itemize}
\item{1.} The interval $r_B^{-1}$ between successive impacts of our pearls
on the earth as estimated from the fact that so far only one
``Tunguska impact''  event has been observed a hundred years ago,
except perhaps the Sodom and Gomorrah event in biblical times.

\item{2.} Just the inverse of $r_B^{-1}$, i.e. $ r_B$.

\item{3.} The dark matter mass density $\rho_{halo}$ in the halo
in the neighborhood of our sun on a kpc scale but away from
us on a scale of the order of the solar system. It is such densities, that
determine the influence of the dark matter on the motion of the stars
and galaxies and it is thus an astronomically measured quantity.
We use it together with the rate $r_b$ to determine - after minor
corrections using the speed $v$ of the pearls  - the average or median
mass $m_B$
of the pearls.

 \item{4.} Using $\rho_{halo}$ as input we estimate
the mass density of dark matter in the solar system near the earth
to be $\approx 2\rho_{halo}$.

\item{5.} Typical speed of the pearl in the region of the
earth, where about half the pearls are supposed to be
linked to the solar system and thus having lower speed, while
about half come from the far out regions of the galactic halo.
This speed is of relevance for determining how often a pearl
hits the earth and thus for how to get the mass by means of the
rate of impacts $r_B $.

\item{6.} The mass $m_B$ of a single pearl determined
from the estimates (1., 4., 5.) above. (This was already done in our
earlier work \cite{Tunguska}.)

\item{7.} The kinetic energy $\frac{1}{2}m_Bv^2$ of the pearl responsible
for the impact and release of energy in Tunguska.

\item{8.} The energy observed as the visible explosion in Tunguska $E_{Tunguska}$.
This of course should at least be smaller than the kinetic energy
of the pearl available for making explosion, since an appreciable
part of the energy will be deposited deeply inside the earth.


\item{9.}  Potential shift for a nucleon in passing through the skin of the pearl $\Delta V \approx
135$ MeV.
We presume that the potential felt by a neutron or a proton inside the pearl is
$\Delta V$ lower than outside due to a lower Higgs field inside the pearl.
Since in appendix \ref{smooth} we argue for now
believing that the Higgs field expectation value is zero inside the pearl,
it follows that $\Delta V=
270$ MeV rather than $135$ MeV and so we have put
``(270 MeV)'' in brackets in column 5.


\item{10.} The force per unit length or equivalently the energy per
unit area of the pearl surface/skin is denoted $S$. In column 4 the value has been fitted to
the hypothesis that the typical pearl size is just on the borderline of stability against
the nuclei being spit out.
Whereas in column five, the value obtained from the fit
to the experimental data is given.
In both columns it is the third root of the tension $S^{1/3}$ which is given.

\item{11.} Here we then give the same third root but now estimated from
theoretical considerations about the Higgs field $\phi_H$ and the effective field $\phi_F$ for the bound state
of $6t +6 \bar{t}$ introduced in our work. Basically it means that $S^{1/3}$
is given by dimensional arguments from the Higgs mass and Higgs field expectation value.

We have two different hypotheses about estimating the effective potential
from which the surface tension $S$ is obtained from a soliton.
``Theory 1)'' involves only the present and the condensate vacuum,
while ``theory 2)'' is based on the assumption of yet
one more vacuum phase.

\item{12.} Ball density or pearl density $\rho_B$ is the specific
density of the bulk of the pearl, i.e. simply the ratio of the mass
to the volume.

\item{13.} The radius $R$ of the ball, mainly thought of as the
radius of the skin sphere.

\item{14.} The homolumo gap calculation is the first
main point of the present article where we obtain
the value for the gap between the lowest unoccupied and the highest
occupied electronic orbits. This energy gap gives rise to radiation from the
dark matter with the frequency essentially equal to the homolumo gap.
It is
our first and most important success that, in the fit of column 5,
this homolumo gap turns out to be order of magnitude-wise equal to 3.5 keV.

\item{15.} In this line we just note down the observed
 X-ray frequency of 3.5 keV, supposedly emitted from dark matter.

\item{16.} The released energy $E_S$ stands for the
energy released when two pearls collide and their common surface
contracts so as to have one combined pearl instead of the
previous two. This released energy is estimated as
the fraction of the surface area contracted away multiplied by
the surface tension $S$. It is written relative to the
Einstein energy of the whole pearl $m_Bc^2$.

\item{17.} ``\# 3.5's as if all $\rightarrow$ 3.5'' means the
number of photons of energy 3.5 keV, which could be produced
from the released energy $E_S$ under the perhaps not realistic
assumption that all the energy went into such 3.5 keV photons.
I.e. it is simply $E_S/(3.5\, \hbox{keV})$.

\item{18.} The spreading time $t_{spread}$ for the hot spot produced in the
collision to spread over the whole pearl, so that its surface gets heated
and energy escapes via higher frequencies than just the 3.5 keV line.

\item{19.} The radiation time $t_{radiation}$ is defined as the time
it would take for the released energy $E_S$ to be emitted, if it was all
emitted as black body radiation
at the temperature
$T =3.5$ keV. Since this is what is expected to happen during the
time interval $t_{spread}$ the fraction of radiation sent out as 3.5 keV
radiation is estimated as the ratio $\frac{t_{spread}}{t_{radiation}}$.
Of course the
emitted number of photons with 3.5 keV cannot be bigger than the number estimated
assuming all the energy goes to 3.5 keV's. Thus, if this ratio $\frac{t_{spread}}{t_{radiation}}\ge 1$
we replace the ratio by 1.

The line ``with 1.5'' means that we took the effective temperature for the
amount of 3.5 keV radiation
emitted to be 1.5 * 3.5 keV instead of just 3.5 keV (see (\ref{1c5})).

\item{20.} The ratio  $\frac{t_{spread}}{t_{radiation}}$ relevant for the
amount of radiation 3.5 keV emitted. In column 5 this ratio is greater than one
and has to be replaced by unity (denoted by ``$2.2 \rightarrow 1$").

The line ``with 1.5'' means that we took the effective temperature for the
amount of 3.5 keV radiation
emitted to be 1.5 * 3.5 keV instead of just 3.5 keV (see (\ref{1c5})).

\item{21.} ``\# 3.5's'' then means the estimate of how
many 3.5 keV photons are truly produced in one collision.
The correction is that only the fraction $\min \{1, \frac{t_{spread}}{t_{radiation}} \}$
of the item 17.: ``\# 3.5's as if all $\rightarrow$ 3.5'' comes out as
3.5 keV radiation

The line ``with 1.5'' means that we took the effective temperature for the
amount of 3.5 keV radiation
emitted to be 1.5 * 3.5 keV instead of just 3.5 keV (see (\ref{1c5})).

 \item{22.} The cross section $\sigma= \pi (2R)^2 $ for the pearls colliding is
supposed to be the geometrical cross section just given by
the radii of the pearls colliding. It is of course $\pi$ times
the square of the sum of the radii of the two pearls.

\item{23.} ``cross section per $\gamma$ (all $\rightarrow $ 3.5)'' formally means
the cross section that
a pearl should have for hitting another pearl if
only one photon (with energy 3.5 keV) was produced per collision and
under the assumption that all energy goes to the 3.5 keV line.

\item{24.}  ``cross section per $\gamma$ (with time ratio)'' formally means the
cross section needed for the collision, if we have to have again
one collision for each photon emitted, but this time taking the
more realistic amount of 3.5 keV photons by them only being produced
during the time $t_{spread}$.

\item{25.} Here we simply divide the ``cross section per $\gamma$
(all $\rightarrow $ 3.5)'' by the mass square of the pearl $M^2 =m_B^2$.

\item{26.} Similarly we divide the number ``cross section per $\gamma$
(with time ratio)'' by $M^2=m_B^2$. This is now the quantity,
which determines the rate of 3.5 keV radiation from various objects,
provided one can estimate the square of the density of dark matter in
those astronomical objects.

The line ``with 1.5'' means that we took the effective temperature for the
amount of 3.5 keV radiation
emitted to be 1.5 * 3.5 keV instead of just 3.5 keV (see (\ref{1c5})).

\item{27.} The quantity $\frac{N\sigma}{M^2}$ extracted from the observations
by Cline and Frey \cite{FreyCline}. This
number should be considered as the experimental value
corresponding to the theoretical value in item 25 or item 26.

\item{28.} The ratio of the actual radius to the ``critical'' radius of
the pearls (our parameter $\xi$), assuming
$\Delta V = 135$ MeV.
The row 28b denotes the value under the
assumption that $\Delta V=
270$ MeV, as will happen for
zero Higgs field $\phi_H$ in the condensate vacuum.

\item{29.} A theoretical estimate of what this
$\xi$ ratio should be provides $\xi =  2^{4/9}\sqrt{4\pi}$,
 which we actually use as a theoretical restriction in
 some of our fits.

\item{30.} This is our estimated value for the heat conductivity $k$.
\end{itemize}

\section{Outlook.}

Now that we
have such a promising model for the dark
matter we should of course
study what our model, with its
parameters getting more and more fixed,
will predict for other phenomena suspected
to come from dark matter such as:
\begin{itemize}
\item {\bf The positron excess:}
Actually our pearls in the situation when
even the surface is very hot - something
happening shortly after the collision -
will emit a lot of electrons, which sit
more loosely than the nucleons. So
presumably a turbulent plasma with
strong fields (electric and/or magnetic)
will appear around the exploding pair of pearls.
This ``little supernova remnant''
could easily be imagined to send out
all sorts of cosmic radiation, including
positrons.
\item{\bf Broad spectrum gamma-rays:}
Although we hope for an of order unity
part  of the
energy coming out as the 3.5 keV X-ray line,
it would of course be almost impossible
that there should not come some radiation
of other frequencies too.

It may be hard to know if such radiation
really comes from dark matter. A mark
of our model should be that the radiation, like
that from annihilation, rather goes
proportional to the square of the dark
matter density than just proportional
to the density itself.
\item {\bf The supernova remnant:}
We really in the near future should estimate
the amount of 3.5 keV radiation that
could appear from a supernova remnant
because of being energized by the
cosmic rays in this remnant rather than
by our collision of pearls, which has no
 reason to be especially frequent in
supernova remnants.
(Remember that indeed such radiation
from a supernova remnant has been observed \cite{Jeltema}.)
\end{itemize}

\appendix
\section{Appendix, Higgs field}\label{RAA}
\subsection{Smooth realization of MPP}\label{smooth}
In previous work on this model we made the assumption that in the
condensate vacuum the Higgs field vacuum expectation value was
decreased compared to the value in the present vacuum by some
factor of order unity, which we in the calculations took to be
a factor 1/2. Actually some estimate in the appendices of an earlier
paper \cite{Tunguska} happened to give us that the Higgs field expectation
value was indeed close to 1/2 of the one in the present vacuum.

In this appendix we shall argue that it is
more natural to take the Higgs field expectation value
in the condensate phase/vacuum to actually be zero.

This argument is based on the hypothesis that there should be a minimal amount
of fine tuning to implement MPP. In fact we require the effective potential, expressed
in terms of the Higgs field $\phi_H$ and an effective field $\phi_F$
for the bound state  of six top and six anti-top
quarks $F$ making the condensate, should be given as a polynomial of lowest
possible order.
Since weak isospin symmetry ensures that
the effective potential
can only depend on the Higgs field through its square $|\phi_H|^2$, we may
think of this effective potential as being a function
 $V_{eff}(|\phi_H|^2,\phi_F)$.

For simplicity we shall only show how the expectation values in the
MPP model tend to go to the boundary under an extra assumption. Namely we assume
that even for the bound state $F$ the field $\phi_F$
only occurs, approximately at least, in the combination $\phi_F^2$ (since the bound state $F$ is its own
antiparticle, we suppose its field is ``real`` in the sense that it is Hermitean).
In this case we can consider the effective potential a function
$V_{eff}(|\phi_H|^2,\phi_F^2)$ and by ``renormalizability `` it would be
required to at most second order in these squares $|\phi_H|^2$ and $\phi_F^2$.

In order to make this effective potential be
as smooth as possible we now postulate that, in terms of these variables $|\phi_H|^2$ and $\phi_F^2$, we can
approximate it by an as low order polynomial as possible.

One should have in mind of course that the variable $|\phi_H|^2$ as well as $\phi_F^2$ must
always be positive or zero
\begin{equation}
 |\phi_H^2|\ge 0, \hbox{ and } \phi_F^2 \ge 0.
\end{equation}
This opens up the possibility for a minimum in this effective potential
 $V_{eff}(|\phi_H|^2,\phi_F^2)$ to occur on the boundary of the allowed region,
i.e. where $|\phi_H|^2$=$0$ or where $\phi_F^2=0$. Then it is no longer needed for there to be
zero derivative w.r.t. both variables. One derivative being zero would do.

One easily sees that if one does not make use of this option of having
a minimum on the boundary of the allowed region, then drawing a
straight line through two degenerate minima as required by MPP would imply
that the effective potential restricted to this line would have two
degenerate minima on the line. Two (degenerate) minima on a line enforces a maximum in between
and the derivative of the effective potential restricted to the line
to have three separate zeros and thus be at least a third order polynomial
on this line. So  without using the possibility of a minimum on the boundary, the restriction
to the line and thus even more the whole effective potential would have to
be at least of fourth order as a function of the variables $|\phi_H|^2$ and $\phi_F^2$.
If we, however, use the option of having one minimum - actually the
one corresponding to the `` condensate vacuum'' - on the border where
 $|\phi_H|^2$=$0$, then the derivative only needs two zeroes. Thus the
effective potential itself $V_{eff}(|\phi_H|^2,\phi_F^2)$ restricted to
the line drawn through the two MPP-minima would only need to be at
most of third order in the variables  $|\phi_H|^2$, and $\phi_F^2$.
But if we let both of the MPP-minima be on border lines then we need
only one maximum of the derivative of the potential restricted to the line.
This would allow the potential itself to be only of second order in the
two squares. This realization of MPP by two minima on the boundaries
is the only way to realize two minima with a dimensionality-wise
renormalizable Lagrangian.

Thus we see that requiring the minimal order of the Taylor expansion
approximation for the effective potential leads to letting both
minima be at a border. That is to say, this smoothness requirement
leads to the Higgs field expectation value in the ``condensate vacuum''
being zero, and the expectation value of the bound state field $\phi_F$ being zero in the present
vacuum. Physically of course this means that the
``condensate vacuum'' has no weak hypercharge nor weak isospin
spontaneous breaking. Especially the quarks are therefore massless in this
vacuum-phase.

In earlier work we could only assume that the Higgs field expectation value
in the condensate vacuum was smaller than in the present vacuum and of order unity
compared to the latter. Now with the expectation that the Higgs field in the condensate vacuum is zero,
in principle we obtain a more accurate prediction for the nucleon potential difference $\Delta V$ on
passing the border of our pearls.
Using the Higgs coupling $g_{HNN}= 1.1 *10^{-3}$ from \cite{gNN}
and making the assumption, as in previous work \cite{Tunguska},
that the Higgs field in the condensate vacuum was
just half of that in the present vacuum $\phi_H = 246$ GeV,
we get the estimate $\Delta V = 135 $ MeV.
So with the zero Higgs field instead, we will get
\begin{equation}
 \Delta V =
270 \, \hbox{MeV}.
\end{equation}

\subsection{An MPP-relation.}

We write our polynomial ansatz for the effective potential
$V_{eff}(|\phi_H|^2,\phi_F^2)$ with both $\phi_H$ and $\phi_F$
occurring only to even powers in the explicit form
\begin{eqnarray}
 V_{eff}(|\phi_H|^2,\phi_F^2)&=& V_H(|\phi_H|^2) + V_F(\phi_F^2)+\lambda_{mix}|\phi_H|^2 \phi_F^2\label{ansatz}\\
\hbox{where } V_H(|\phi_H|^2)&=& \lambda_H (|\phi_H|^2 - v_H^2)^2 \label{VH}\\
\hbox{and } V_F(\phi_F^2)&=& \lambda_F (\phi_F^2 - v_F^2)^2\label{VF}
\end{eqnarray}
In the model suggested above the degenerate minima occur on the
axes where $\phi_H=0$ and $\phi_F=0$ respectively. Hence the term
$\lambda_{mix}|\phi_H|^2 \phi_F^2$ representing the interaction
between the two fields is zero at these minima. The energy density
for the two minima are easily seen to be
\begin{eqnarray}
 V_{eff}(\hbox{``min at present vacuum''})&=& V_{eff}(v_H^2,0) = \lambda_Fv_F^4
\end{eqnarray}
and
\begin{eqnarray}
V_{eff}(\hbox{``min at condensate vacuum''})&=& V_{eff}(0,v_F^2) = \lambda_Hv_H^4;
\end{eqnarray}
respectively. So the multiple point principle requirement that these two minima be equally deep
means that we must have
\begin{eqnarray}
 \lambda_Fv_F^4&=&\lambda_Hv_H^4.\label{MPPrelfirst}
\end{eqnarray}
Denoting the mass of the $F$-particle in the condensate vacuum
by $m_F|_{condensate}$ we have
\begin{eqnarray}
 m_F|_{condensate}^2 &=& 8\lambda_Fv_F^2\label{mF}.
\end{eqnarray}
Similarly the Higgs mass in the present vacuum is
\begin{eqnarray}
 m_H|_{present}^2 &=& 4\lambda_Hv_H^2,\label{mH}.
\end{eqnarray}
Thus our just derived MPP-relation becomes
\begin{eqnarray}
 m_F|_{condensate}v_F &=& \sqrt{2} m_H|_{present} v_H\label{MPPrel}.
\end{eqnarray}
This equation is interesting because it tells that if the
bound state $F$ is heavy in the condensate phase compared to the Higgs mass, then
its vacuum expectation value in the condensate phase
will have to be correspondingly small.

\subsection{Minimal $\lambda_{mix}$}

In order to ensure that the assumed two degenerate minima are indeed the
lowest energy states of the system, we have to require that the
effective potential does not fall to an even lower value than these minima
in some other place $(\phi_H,\phi_F)$. If we did not have
the $\lambda_{mix} |\phi_H|^2\phi_F^2$ term, there would indeed be a
minimum deeper than the two minima at the borders, which would actually
not be true minima themselves anymore but only saddle points in
$V_{eff}$. In this case, with $\lambda_{mix} = 0$, the true minimum would be of course
be at the point $(\phi_H,\phi_F) = (v_H,v_F)$. In order to fill up this
minimum so that the value of $V_{eff}$ becomes at least $\lambda_Hv_H^4 =\lambda_Fv_F^4$, we need
the mixing term to satisfy
\begin{eqnarray}
\lambda_{mix}v_H^2v_F^2 &\ge & \lambda_Hv_H^4 =\lambda_Fv_F^4\\
\Rightarrow \quad \lambda_{mix} &\ge& \sqrt{\lambda_H\lambda_F}.
\end{eqnarray}
But we also have to make the slope of $V_{eff}$ at the present vacuum minimum
be non-negative in the increasing $\phi_F$ direction. Since
near the present vacuum the dominant
term from $V_F$ is $-2\lambda_Fv_F^2 \phi_F^2$, we in fact require
\begin{eqnarray}
 \lambda_{mix}v_H^2\phi_F^2&\ge& 2\lambda_Fv_F^2 \phi_F^2\\
\Rightarrow \quad \lambda_{mix}v_H^2&\ge& 2\lambda_Fv_F^2
\end{eqnarray}
and analogously:
\begin{eqnarray}
 \lambda_{mix}v_F^2&\ge& 2\lambda_Hv_H^2.
\end{eqnarray}
Then multiplication gives:
\begin{eqnarray}
\lambda_{mix}^2 v_H^2 v_F^2 &\ge& 4\lambda_H\lambda_F v_H^2 v_F^2\\
\Rightarrow \quad \lambda_{mix}&\ge& 2\sqrt{\lambda_H\lambda_F}.\label{lambdamixgesqrt}
\end{eqnarray}

We now show that the inequality (\ref{lambdamixgesqrt}) ensures that the effective
potential is higher than or equal to the two hoped for MPP-minima, by writing it in the form
\begin{eqnarray}
 &&V_{eff}(|\phi_H|^2,\phi_F^2)-\lambda_Hv_H^4 -\lambda_Fv_F^4\nonumber\\
&=& (\lambda_{mix} - 2\sqrt{\lambda_H\lambda_F})|\phi_H|^2\phi_F^2
+ (\sqrt{\lambda_H}|\phi_H|^2 +\sqrt{\lambda_F}\phi_F^2)^2 +m_{H \; tac}^2|\phi_H|^2 +\frac{1}{2}m_{F \; tac}^2\phi_F^2\\
&=& (\lambda_{mix} - 2\sqrt{\lambda_H\lambda_F})|\phi_H|^2\phi_F^2
+ (\sqrt{\lambda_H}|\phi_H|^2 +\sqrt{\lambda_F}\phi_F^2)^2 -2\lambda_Hv_H^2|\phi_H|^2 -2\lambda_Fv_F^2\phi_F^2\\
&=& (\lambda_{mix} - 2\sqrt{\lambda_H\lambda_F})|\phi_H|^2\phi_F^2
+ (\sqrt{\lambda_H}|\phi_H|^2 +\sqrt{\lambda_F}\phi_F^2)^2 -2\sqrt{\lambda_F}v_F^2(\sqrt{\lambda_H}|\phi_H|^2 +\sqrt{\lambda_F}\phi_F^2)\nonumber\\
&=& (\lambda_{mix} - 2\sqrt{\lambda_H\lambda_F})|\phi_H|^2\phi_F^2
+ (\sqrt{\lambda_H}(|\phi_H|^2-v_H^2) +\sqrt{\lambda_F}\phi_F^2)^2 -\lambda_Hv_H^4\\
&=& (\lambda_{mix} - 2\sqrt{\lambda_H\lambda_F})|\phi_H|^2\phi_F^2
+ (\sqrt{\lambda_H}|\phi_H|^2 +\sqrt{\lambda_F}(\phi_F^2-v_F^2))^2 -\lambda_Hv_H^4
\end{eqnarray}
Here we had to have in mind equation (\ref{MPPrelfirst}) under the assumption of our MPP.
This form of the effective potential contains
\begin{itemize}
 \item the term $(\lambda_{mix} - 2\sqrt{\lambda_H\lambda_F})|\phi_H|^2\phi_F^2$, which
is non-negative for $\lambda_{mix}\ge 2\sqrt{\lambda_H\lambda_F}$.
\item The total square term that can be written both as
$(\sqrt{\lambda_H}|\phi_H|^2 +\sqrt{\lambda_F}(\phi_F^2-v_F^2))^2$
and as $(\sqrt{\lambda_H}(|\phi_H|^2-v_H^2) +\sqrt{\lambda_F}\phi_F^2)^2 $,
which is zero in the minima corresponding to the present and condensate vacua.
\item Finally there is the not so important cosmological constant term,
which in $V_{eff}$ is
$\lambda_H v_H^4 = \lambda_Fv_F^4$.
\end{itemize}
So from this form one sees that
\begin{eqnarray}
 V_{eff}(|\phi_H|^2, \phi_F)&\ge&  (\lambda_{mix} - 2\sqrt{\lambda_H\lambda_F})|\phi_H|^2\phi_F^2 +\lambda_Hv_H^4
\end{eqnarray}
and thus
\begin{eqnarray}
 V_{eff}(|\phi_H|^2, \phi_F)&\ge&  V_{eff}(v_H^2,0)= V_{eff}(0,v_F^2)\ \nonumber
\end{eqnarray}
provided that
\begin{eqnarray}
 \lambda_{mix}&\ge& 2\sqrt{\lambda_H\lambda_F}.
\end{eqnarray}

In the case when we have equality in the inequality
\begin{eqnarray}
 \lambda_{mix}&\ge& 2\sqrt{\lambda_H\lambda_F}
\end{eqnarray}
the effective potential takes the form
\begin{eqnarray}
 &&V_{eff}(|\phi_H|^2 , \phi_F^2)\nonumber\\
&=&
(\sqrt{\lambda_H}|\phi_H|^2 + \sqrt{\lambda_F}\phi_F^2)^2+
m_{H \; tac}^2 |\phi_H|^2 + \frac{1}{2}m_{F \; tac}^2\phi_F^2.
\end{eqnarray}
Here the masses in the Lagrangian are written in the usual form and
carry the index $tac$ standing for ``tachyonic''. They are given by the expressions
\begin{eqnarray}
m_{H \; tac}^2 &=& -2\lambda_Hv_H^2
\end{eqnarray}
and
\begin{eqnarray}
 m_{F \; tac}^2 &=& -4\lambda_Fv_F^2.
\end{eqnarray}
In fact these tachyonic masses are proportional to the masses of the
bound state $F$ (\ref{mF}) and the Higgs $H$ (\ref{mH}) in the
``condensate'' phase and the ``present'' phase respectively.
If we now impose the MPP-relation (\ref{MPPrel})
we get that the combination of mass terms
only depends on the quantity $\sqrt{\lambda_H}|\phi_H|^2 +\sqrt{\lambda_F}\phi_F^2$,
in terms of which we have written the fourth order part.
In other words with MPP and $\lambda_{mix}$ taken minimal, the whole
effective potential only depends on the quantity
$\sqrt{\lambda_H}|\phi_H|^2 +\sqrt{\lambda_F}\phi_F^2$.
So under these two assumptions the effective potential
is constant along contour curves  for this quantity.
Especially one such contour curve connects the two
degenerate minima. This then implies that a field can
vary smoothly along this curve from one minimum to the
other one. And that in turn implies that a soliton solution
giving zero tension $S$ can be found in this
special case with minimum $\lambda_{mix}$.


\subsection{The Surface tension $S$, Introduction.}

We already found above that, for the coupling constant
$\lambda_{mix}$ taking its minimally allowed value
$2\sqrt{\lambda_H\lambda_F}$, there is a flat direction
from the one minimum to the other one. Thus the
surface tension for the wall separating the ``present''
and the ``condensate'' vacua would be $S=0$ in this
minimum case.

Increasing $\lambda_{mix}$ from this minimum value, the surface tension $S$ will
of course increase. But, however high
$\lambda_{mix}$ would be, it can never lead to a higher $S$ than the
value obtained when the route of the field combination $(\phi_H,\phi_F)$
in the soliton, as one passes the wall, only goes through combinations
wherein one of the two fields is zero. In fact one could consider a
potential soliton having the field combination following the axes
in the $(\phi_H,\phi_F)$-plane from the one minimum via the
origin in the field combination space to the other minimum.
But if $\lambda_{mix}$
is not too large and positive,
then there is the possibility of finding a ``lower'' path
for the soliton solution, which would give a lower tension
$S$. Thus the tension estimated taking the path of the soliton to be along the axes
would become an upper limit for $S$.

In order to estimate the size of this upper limit
we could consider a couple of helpful auxiliary problems:

Formally we can take a single field theory, say the Higgs theory
with $V_H(\phi_H) = \lambda_H (\phi_H^2-v_H^2)^2$ as if the
Higgs field was real but could be both positive and negative.
We would then have a model with the symmetry
$\phi_H \rightarrow -\phi_H$, and two degenerate vacua because of this
symmetry. (We would have MPP by this symmetry).
We can now calculate the energy per unit area
of a soliton in which the field $\phi_H$ goes from one of these
minima to the other one. We call the energy per unit area  or the tension
of such a wall $S_H$. For the analogous problem, obtained by using the $\phi_F$
field and the $V_F(\phi_F)$ potential given by (\ref{VF}) and
letting the soliton field go from $v_F$ to $-v_F$, we call the
tension $S_F$. Then the tension for the soliton field going via the
origin $(0,0)$ would be
\begin{eqnarray}
 S&=&\frac{1}{2} (S_H +S_F).
\end{eqnarray}
The factor $\frac{1}{2}$ comes in because, in going from the one minimum to the other one
in the full model with two fields via the origin, one
only goes half of the distance one goes in
our two auxiliary models.

\subsection{Calculating the tension $S$.}\label{theory1}

In this subsection we shall first compute the tension or energy per unit area
for a solitonic transition from a region with $\phi_H =-v_H$ to
one with $\phi_H=v_H$ with the potential $V_H(\phi_H)= \lambda_{H}(\phi_H^2 -v_H^2)^2$.
Since the potential energy density at the start and end points is just
zero, the equation for the derivative of the field $\phi_H$ with respect
to the coordinate $x$ in the direction perpendicular to the surface between the two phases
is
\begin{eqnarray}
 \left ( \frac{\partial \phi_H}{\partial x} \right )^2 &=& V_H(\phi_H) = \lambda_H(\phi_H^2 - v_H^2)^2. \\
\end{eqnarray}
The first step is to find $\phi_H$ as a function of the perpendicular coordinate $x$.
Taking the square root etc. of this equation gives
\begin{eqnarray}
 (\pm) \frac{\partial \phi_H}{\partial x} &=& \sqrt{\lambda_H}(v_H^2 -\phi_H^2)\\
\Rightarrow \quad \frac{d \phi_H}{\sqrt{\lambda_H}(v_H^2 -\phi_H^2)}&=& dx\\
\Rightarrow \quad \tanh^{-1}\left(\frac{\phi_H}{v_H}\right) &=&\sqrt{\lambda_H}v_Hx
\end{eqnarray}
with the appropriate choice of origin. Thus one obtains
\begin{eqnarray}
 \phi_H &=& v_H \tanh(\sqrt{\lambda_H}v_Hx )
\end{eqnarray}

The energy per unit area of the wall is then
\begin{eqnarray}
 S_H &=&\int_{\infty}^{\infty} 2 V_{H \; eff}dx\\
&=&  \int_{\infty}^{\infty} 2 \lambda_H (v_H^2 - \phi_H^2)^2 dx\\
&=&  \int_{\infty}^{\infty} 2 \lambda_H (v_H^2 - v_H^2 \tanh^2(\sqrt{\lambda_H } v_H x) )^2 dx\\
&=&2\sqrt{\lambda_H}v_H^3\int_{\infty}^{\infty} \frac{1}{\cosh^4(u)} du\\
&=&\frac{8\sqrt{\lambda_H}v_H^3}{3}
\end{eqnarray}
Using (\ref{mH}) this means
\begin{eqnarray}
 S_H &=& \frac{4m_Hv_H^2}{3}.
\end{eqnarray}

Similarly we find
\begin{eqnarray}
 S_F &=& \frac{2m_Fv_F^2}{3} = \frac{m_H}{m_F}S_H \approx \frac{1}{6}S_H,
\end{eqnarray}
where we have taken $m_F \sim 750$ GeV.

So the upper limit on the tension becomes
\begin{eqnarray}
 S&=& \frac{1}{2} (S_H + S_F) \approx \frac{7}{12} S_H=  \frac{7}{9}m_Hv_H^2,
\end{eqnarray}
and the upper limit for $S^{\frac{1}{3}}$ is
\begin{eqnarray}
 S^{\frac{1}{3}}&=&140\, \hbox{GeV}.
\end{eqnarray}
Hence the true tension $S$ must satisfy
\begin{eqnarray}
 S^{\frac{1}{3}} \le 140\, \hbox{GeV}.\label{Slimit}
\end{eqnarray}

\subsection{Rescue Attempt}\label{RA}
There is one rather natural story that can provide a better fit
with the theoretical cubic root of the surface tension $S^{1/3}$ not deviating so
much from the fitted value.

This proposal is based on saying that, since we anyway have assumed the multiple point principle, we can easily expect
even more than two degenerate vacua (namely the present vacuum and the
condensate vacuum) at low energy.
Although we do not have any explicit third vacuum in mind, we could just abstractly
claim that there could exist another vacuum degenerate with the other two vacua.
With say three vacua at a low energy scale, it is natural to consider that we have
three effective fields $\phi_H$, $\phi_F$ and $\phi_{F'}$ instead of only the first two
as we used above. Also we shall keep our assumption from appendix \ref{smooth} that
we approximate the effective potential $V_{eff}$ by a polynomial only up to fourth order terms in the fields
 $\phi_H$, $\phi_F$ and $\phi_{F'}$ and only with even powers of these fields. Then
 the effective potential would be a second order polynomial in the three square variables
 $|\phi_H|^2$, $\phi_F^2$ and $\phi_{F'}^2$. Now to have three degenerate minima - as
 required by MPP with three vacua - this polynomial in the three variables has to have
 three equally deep minima.

We shall now
 see that having the three assumed degenerate minima leads to the effective potential
 being completely constant/flat along the planar surface in the squared fields space spanned
 by the positions of these three minima.
We first remark that along any line in the squared field space
 the effective potential is still of course also a polynomial of the second order
 in the squared fields. But now, if you consider the line spanned by two of the minimum
 positions, the effective potential restricted to such a line would have to have two (degenerate) minima. It follows
 that the derivative of the effective potential $\frac{dV_{eff}}{dl}$ (where $l$ parameterizes this line)
 would have to have at least two zeros, namely one at each of the minimum-positions.
 The effective potential itself  would even have to have a maximum in between these
 two minima and thus the derivative would have at least three zeros. But the existence of three zeros
 is impossible for a first order polynomial - the derivative is only first order - on a line except if the
 polynomial is totally zero. By this argument there have to be flat directions for the
 effective potential along all lines passing through just two of the minima positions.

 Let us consider the set of second derivatives of the effective potential
 $\frac{\partial^2V_{eff}}{\partial \phi_i\partial\phi_j}$ forming a (3$\times$3) matrix.
 We see, from the just derived flat directions, that this matrix taken at one of the minimum points
 must have a zero eigenvalue corresponding to each of the flat directions
 extending to the other two minima. But having now two
 zero eigenvalues the whole plane spanned by the two eigenvectors would give us
 a flat direction, i.e. a whole flat plane along which the effective potential
 would be degenerate with the MPP-minima.

 Note that, in the here described approximation, the surface tensions $S$ for domain walls
 separating any two of the three phases/vacua would be zero. So only if the
 effective potential is after all not a second order polynomial in the square fields
 but e.g. of third order (meaning of sixth order in the fields themselves) would we obtain
 non-zero tensions.

 This means that, to the degree that our approximation was a good one,
 the tensions would be appreciably smaller than first estimated, which
 could greatly improve the quality of our fit.

 \subsection{Order of magnitude for the Tension $S$} \label{theory2}

 If we can give a dimensional argument for the degree of suppression of the
 sixth order terms compared to the fourth order or lower terms, we could deliver
 an order of magnitude estimate for the tension $S$ for the domain walls between the phases.

 Let us adopt the philosophy that we are concerned with phases lying at a scale of the Higgs mass
 $M_H = 125\ \hbox{GeV}$ while the physics relevant for the effective couplings $\lambda_H, \lambda_F,\lambda_{F'}, m_F^2 ,...$
 is rather the physics scale of the bound state $m_F$, which we have previously estimated
 to have a mass of order 750 GeV. If so we could claim that the Taylor expansion, which we terminated
 with a fourth order polynomial above, effectively has its terms go down by a factor
 $\left ( \frac{125}{750}\right )^2 =\frac{1}{36} =0.028$ for each pair of mass dimensions.
 So, for example, the sixth order terms that alone contribute to the tensions in the present three
 vacuum model will be reduced compared to the typical order of magnitude expected had the up to
 fourth order terms indeed contributed - as happens with only two vacua - by a factor of
 36. In the two degenerate vacua case we obtained an $S^{1/3}$-upper limit (\ref{Slimit}) of
the order 140 GeV or a typical value say of 100 GeV. This value would go down by a factor of
 $\sqrt[3]{36} = 3.3$, meaning to $S^{1/3}$ = (100 GeV)/3.3 = 30 GeV.
%



\end{document}